\documentclass[12pt, draftclsnofoot, onecolumn]{IEEEtran}
\usepackage{amsmath}
\usepackage{amssymb}
\usepackage{amsthm}
\usepackage{cite}
\usepackage{color}
\usepackage{xcolor}
\usepackage{caption}
\usepackage{epsfig,latexsym}
\usepackage{float}
\usepackage{fancyhdr}
\usepackage{graphicx}
\usepackage{indentfirst}
\usepackage{mdwmath}
\usepackage{mdwtab}
\usepackage{subfigure}
\usepackage{setspace}
\usepackage{times}
\usepackage{url}
\usepackage{multirow}

\newtheorem{remark}{Remark}
\newtheorem{theorem}{Theorem}

\newtheorem{lemma}{Lemma}

\newtheorem{corollary}{Corollary}

\newtheorem{proposition}{Proposition}

\begin{document}

\title{Reconfigurable Intelligent Surface Aided NOMA Networks}
\author{Tianwei Hou,~\IEEEmembership{Student Member,~IEEE,}
        Yuanwei Liu,~\IEEEmembership{Senior Member,~IEEE,}
        Zhengyu Song,
        Xin Sun,
        Yue Chen,~\IEEEmembership{Senior Member,~IEEE,}\\
        and Lajos Hanzo,~\IEEEmembership{Fellow,~IEEE}

\thanks{This work is supported by the Beijing Natural Science Foundation under Grant 4194087.} 
\thanks{T. Hou, Z. Song and X. Sun are with the School of Electronic and Information Engineering, Beijing Jiaotong University, Beijing 100044, China (email: 16111019@bjtu.edu.cn, songzy@bjtu.edu.cn, xsun@bjtu.edu.cn).}
\thanks{Y. Liu and Yue Chen are with School of Electronic Engineering and Computer Science, Queen Mary University of London, London E1 4NS, U.K. (e-mail: yuanwei.liu@qmul.ac.uk, yue.chen@qmul.ac.uk).}
\thanks{L. Hanzo is with University of Southampton, Southampton, U.K. (email:lh@ecs.soton.ac.uk).}
}

\maketitle

\begin{abstract}
Reconfigurable intelligent surfaces (RISs) constitute a promising performance enhancement for next-generation (NG) wireless networks in terms of enhancing both their spectrum efficiency (SE) and energy efficiency (EE). We conceive a system for serving paired power-domain non-orthogonal multiple access (NOMA) users by designing the passive beamforming weights at the RISs. In an effort to evaluate the network performance, we first derive the best-case and worst-case of new channel statistics for characterizing the effective channel gains. Then, we derive the best-case and worst-case of our closed-form expressions derived both for the outage probability and for the ergodic rate of the prioritized user. For gleaning further insights, we investigate both the diversity orders of the outage probability and the high-signal-to-noise (SNR) slopes of the ergodic rate. We also derive both the SE and EE of the proposed network. Our analytical results demonstrate that the base station (BS)-user links have almost no impact on the diversity orders attained when the number of RISs is high enough. Numerical results are provided for confirming that: i) the high-SNR slope of the RIS-aided network is one; ii) the proposed RIS-aided NOMA network has superior network performance compared to its orthogonal counterpart.
\end{abstract}


\section{Introduction}

The demand for next-generation (NG) networks having high energy efficiency (EE) has been rapidly increasing~\cite{5G_NR}. A variety of sophisticated wireless technologies have been proposed for NG networks, including massive multiple-input multiple-output (MIMO) and millimeter wave (mmWave) communications~\cite{5G_NR_2}. Recently, cost-efficient reconfigurable intelligent surfaces (RISs) have been proposed for cooperative NG networks~\cite{LIS_zhangjiayi_mag,LIS_smart,RIS_mag_basar}.

To enhance both the spectrum efficiency (SE) and EE of NG networks, non-orthogonal multiple access (NOMA) has been proposed as a promising technique of opportunistically capitalizing on the users' specific channel state information (CSI) differences~\cite{NOMA_mag_Ding,PairingDING2016,Massive_NOMA_Cellular_IoT}. NOMA networks are capable of serving multiple users at different quality-of-service (QoS) requirements in the same time/frequency/code resource block~\cite{NOMA_5G_beyond_Liu,Islam_NOMA_survey,NOMA_large_heter}. Hence, inspired by the aforementioned benefits of NOMA and RIS techniques, we explore the network's performance enhanced by the intrinsic integration of the power-domain NOMA\footnote{In this article, we use NOMA to refer to power-domain NOMA for simplicity.} and RIS techniques in the downlink (DL).

\subsection{Prior Work}

In recent years, RIS based techniques have received considerable attention owing to their beneficial applications~\cite{LIS_magazine_multi_scenarios,reconfig_meta_surf_1,reconfig_meta_surf_2}. The RIS aided system comprises an array of intelligent surface units, each of which can independently absorb energy and shift the phase of the incident signal. By appropriately adjusting the reflection angles and amplitude coefficients of RIS elements, the electromagnetic signal can be reconfigured for wireless transmission.
The performance of RIS-aided and relay-assisted networks was compared in~\cite{LIS_compare_relay}, indicating that RIS-aided networks may have better network performances, provided that the number of RISs is high enough. The associated energy consumption model was proposed in~\cite{glob_energy_model,energy_model_LIS}, where the EE of the proposed network was optimized. Numerous application scenarios, such as RISs aided physical layer security relying on cooperative jamming techniques have also been considered~\cite{Renzo_PHY_security_confe,PLS_LIS_ZhangRui}. The RIS components are capable of blocking the signal of eavesdroppers, hence enhancing the secrecy performance. RIS assisted simultaneous wireless information and power transfer (SWIPT) was proposed in~\cite{Swipt_LIS_ZhangRui} for the users located in coverage-holes. In the 5G new radio (NR) standard, the coverage area is significantly reduced for carriers beyond 6GHz~\cite{Renzo_mmwave_signal_enhancement}. Hence, a sophisticated signal alignment strategy was employed at the RISs for coverage area enhancement in mmWave scenarios~\cite{Lv_coverage_enhancement}.
However, in most previous research, continuous amplitude coefficients and phase shifts were assumed at the RISs~\cite{Zhou_MISO_multi_cluster}, whilst in practice the phase shifts of RISs may not be continuous. Thus discrete phase shifts were considered in~\cite{ZhangRui_MISO_beams_discrete_2} for a multiple-input single-output (MISO) assisted RIS network. The channel capacity of a RIS-aided MIMO network was maximized, where both analog and digital beamforming, as well as hybrid beamforming were considered~\cite{shuowen_RIS}. Furthermore, focusing on the user's fairness, a fairness-oriented design (FOD) was proposed in a RIS-aided MIMO network~\cite{Hou_RIS_MIMO_global_algrithm}.

To further enhance both the SE and EE of the DL, NOMA and RIS techniques were integrated in~\cite{DING_RIS_NOMA_letter}. The RISs can be deployed for enhancing the power level of the cell-edge users, where the cell-center users treat the reflected signal as interference~\cite{DING_RIS_NOMA_letter}. Both continuous and discrete phase shifters were used in a RIS-aided MISO NOMA network~\cite{yuanwei_NOMA_RIS}. Naturally, the BS-user link plays a key role~\cite{MISO_with_directlink}. A RIS-aided NOMA network was also investigated in~\cite{NOMA_RIS_Fu}, whilst the BS-user link and the BS-RIS link, as well as the RIS-user link were assumed to experience Rayleigh fading. The associated bit error ratio (BER) was evaluated in the case of Rayleigh fading in~\cite{LIS_perform_Anal}.
However, both the BS and RISs are part of the infrastructure, and the RISs are typically positioned for exploiting the line-of-sight (LoS) path with respect to the fixed BS in NG networks for increasing the received signal power. Hence, the impact of fading environments on RIS networks has also attracted attention~\cite{RIS_NOMA_Rice}.
A fairness-oriented algorithm was proposed in a RIS-aided NOMA network~\cite{RIS_NOMA_Rice}, where Rician fading channels were used for modelling the channel gains. Note that when the Nakagami and Rice fading parameter obey the following constraint $m=\frac{(K+1)^2}{2K+1}$, these fading channels are identical~\cite[eq. (3.38)]{wireless_communication_goldsmith}.

\subsection{Motivations and Contributions}

The above-mentioned papers mainly studied the network's fairness, whilst there is a paucity of investigations on the SE improvement of NOMA networks. To comprehensively analyze the network's performance enhanced by RISs, a RIS-aided SISO-NOMA network is proposed.
Motivated by the potential joint benefits of RISs and NOMA networks, whilst relying on analog beamforming~\cite{Yi_anlog_beam}, in this article we will analyse the performance of a RIS-aided NOMA DL scenario, where a priority-oriented design (POD) is proposed, which is also capable of enhancing the SE. In the proposed POD, we improve the performance of the user having the best channel gain, where all other users rely on RIS-aided beamforming. In contrast to previous contributions~\cite{Hou_RIS_MIMO_global_algrithm}, we will show that the proposed POD outperforms the FOD in terms of its SE.

Against to above background, our contributions can be summarized as follows:
\begin{itemize}
  \item We propose a novel RIS-aided NOMA network, where a POD is employed for enhancing the SE. The impact of the LoS transmission on the reflected BS-RIS-user links are exploited. Furthermore, the impact of the proposed design on the attainable performance is characterized in terms of its outage probability (OP), ergodic rate, SE and EE.
  \item Explicitly, we derive closed-form expressions of both the OP and of the ergodic rate for the proposed RIS-aided NOMA network. Both the best-case and worst-case of the OP and of the ergodic rate are derived. Both accurate and approximate closed-form expressions are derived. Furthermore, both the diversity orders and high-SNR slopes are obtained based on the OP and ergodic rate. The results confirm that the diversity order can be enhanced by increasing the number of RISs.
  \item The simulation results confirm our analysis, illustrating that: 1) the BS-user link can be ignored when the number of RISs is high enough; 2) the RIS-aided NOMA network relying on the optimal power allocation factors is capable of outperforming its OMA counterpart; 3) the SE of the proposed POD can be significantly enhanced compared to the FOD, when the number of RISs is high enough.
\end{itemize}

\subsection{Organization and Notations}

The rest of the paper is organized as follows. In Section \uppercase\expandafter{\romannumeral2}, the model of RIS-aided NOMA networks is discussed. Our analytical results are presented in Section \uppercase\expandafter{\romannumeral3}, while our numerical results are provided in Section \uppercase\expandafter{\romannumeral4} for verifying our analysis, followed by our conclusions in Section \uppercase\expandafter{\romannumeral5}. 
The distribution of a circularly symmetric complex Gaussian (CSCG) random variable with mean $x$ and covariance matrix $k$ is denoted by $\mathcal{CN} (x,k)$; and $\sim$ stands for ``distributed as''. $\mathbb{P}(\cdot)$ and $\mathbb{E}(\cdot)$ represent the probability and expectation, respectively. 


\section{System Model}

\begin{figure}[t!]
\centering
\includegraphics[width =3in]{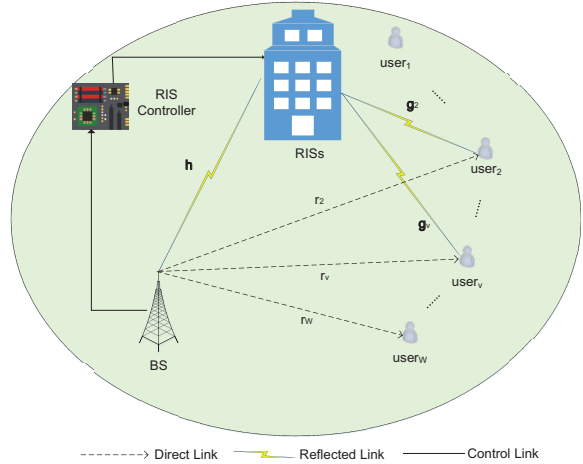}
\caption{Illustration of RIS-aided NOMA networks.}
\label{system_model}
\vspace{-0.2in}
\end{figure}

Let us consider the RIS-aided NOMA DL, where a BS equipped with a single transmit antenna (TA) is communicating with $W$ users, each equipped with a single receive (RA) antenna. We have $N>1$ intelligent surfaces at the appropriate location. By appropriately adjusting the reflection angles and amplitude coefficients of the RIS elements, the electromagnetic signal can be beneficially manipulated. Fig.~\ref{system_model} illustrates the wireless communication model for a single BS.

\subsection{RIS-Aided SISO-NOMA Network}

We first provide a fundamental model to illustrate the network performance affected by RISs.
In order to illustrate the LoS links between the BS and RISs, the small-scale fading vector is defined as
\begin{equation}\label{channel matrix,BS to LIS}
{{\bf{h}}} = \left[ {\begin{array}{*{20}{c}}
{{h_{1}}}\\
 \vdots \\
{{h_{N}}}
\end{array}} \right],
\end{equation}
where ${\rm \bf h}$ is a $(N \times 1)$-element vector whose elements represent the Nakagami fading channel gains.
The probability density function (PDF) of the elements can be expressed as
\begin{equation}\label{channel PDF,eq3}
{f}_1(x) = \frac{{{m_1}^{m_1} {x^{{m_1} - 1}}}}{{\Gamma ({m_1})}}{e^{ - {{{m_1}x}}}},
\end{equation}
where $m_1$ denotes the fading parameter, and $\Gamma ({ \cdot })$ represents the Gamma function. Note that $\Gamma ({m_1})=({m_1}-1)!$ when $m_1$ is an integer.

It is assumed that there are a total of $W$ users in the cluster, where the pair of users, user $W$ and user $v$ with $1 \le v < W$, are superimposed for DL transmission in NOMA.
Hence, the small-scale fading vector between the RISs and user $W$ is defined as
\begin{equation}\label{channel matrix,LIS to w user}
{{\bf{g}}_{W}} = \left[ {\begin{array}{*{20}{c}}
{{g_{W,1}}}& \cdots &{{g_{W,N}}}
\end{array}} \right].
\end{equation}
Similarly, the small-scale fading vector between the RISs and user $v$ is given by
\begin{equation}\label{channel matrix,LIS to v user}
{{\bf{g}}_{v}} = \left[ {\begin{array}{*{20}{c}}
{{g_{v,1}}}& \cdots &{{g_{v,N}}}
\end{array}} \right],
\end{equation}
where ${\rm \bf g}_{W}$ and ${\rm \bf g}_{v}$ are $(1 \times N)$-element vectors whose elements represent the Nakagami fading channel gains having fading parameters of $m_W$ and $m_v$, respectively.

Due to the strong scattering environment, the BS-user link between the BS and user $W$ as well as that between the BS and user $v$ are modelled by Rayleigh fading, which can be expressed as ${r_{W}}$ and ${r_{v}}$, respectively. 

In DL transmission, the BS sends the following signal to the paired NOMA users:
\begin{equation}\label{information bearing}
{\bf{s}} = \alpha _v^2{s_v} + \alpha _W^2{s_W},
\end{equation}
where $s_v$ and $s_W$ denote the signal intended for user $v$ and user $W$, respectively, with $\alpha _v^2$ and $\alpha _W^2$ representing the power allocation factors of user $v$ and user $W$, respectively. Based on the NOMA protocol, $\alpha _v^2+ \alpha _W^2 =1$.

Without loss of generality, we focus our attention on user $W$, and the signal received by user $W$ from the BS through RISs is given by
\begin{equation}\label{received user signal}
{y_w} = \left( {{{\bf{g}}_w}{\bf{\Phi h}}d_1^{ - \frac{{{\alpha _l}}}{2}}d_{2,w}^{ - \frac{{{\alpha _l}}}{2}} + {r_w}d_{3,w}^{ - \frac{{{\alpha _n}}}{2}}} \right)p{\bf{s}} + {N_0},
\end{equation}
where ${p}$ denotes the transmit power of the BS, ${\rm \bf \Phi}  \buildrel \Delta \over = {\rm{diag}}\left[ {{\beta_1} {\phi _1}, {\beta_2}{\phi _2}, \cdots,{\beta_{N}} {\phi _{N}}} \right]$ is a diagonal matrix, which accounts for the effective phase shift applied by all intelligent surfaces, $\beta_n  \in \left( {0,1} \right]$ represents the amplitude reflection coefficient of RISs, while ${\phi _n} = \exp (j{\theta _n}), j=\sqrt{-1}, \forall n = 1,2 \cdots ,N$, and ${\theta _n} \in \left[ {0,2\pi } \right)$ denotes the phase shift introduced by the $n$-th intelligent surface. It is assumed that the CSIs are perfectly known at the RIS controller~\cite{ZhangRui_MISO_beams_discrete_2}. $d_1$ and $d_{2,w}$ denote the distance between the BS and RISs as well as that between the RISs and user $w$, while $d_{3,w}$ denotes the distance between the BS and user $w$. Furthermore, $\alpha_l$ and $\alpha_n$ denote the path loss exponent of the BS-RIS-user links and BS-user links.
Finally, $N_0$ denotes the additive white Gaussian noise (AWGN), which is modeled as a realization of a zero-mean complex circularly symmetric Gaussian variable with variance ${\sigma ^2}$.

\section{RIS Design for the Prioritized User in NOMA Networks}

In this section, we first design the phase shifts and reflection amplitude coefficients for the RISs. Our new channel statistics, OPs, ergodic rates, SE and EE are illustrated in the following subsections.

\subsection{RIS Design}

When the direct BS-user signal and reflected BS-RIS-user signals are co-phased, the channel gain of user $W$ is given by
\begin{equation}\label{maximum achievable gain}
{\left| {{{\tilde h}_W}} \right|^2} = {\left| {{{\bf{g}}_W}{\bf{\Phi h}}d_1^{ - \frac{{{\alpha _l}}}{2}}d_{2,W}^{ - \frac{{{\alpha _l}}}{2}} + {r_W}d_{3,W}^{ - \frac{{{\alpha _n}}}{2}}} \right|^2}.
\end{equation}

It is assumed that there are $W$ users in the cluster, and then the achievable channel gains of users $1 \cdots W$ are ordered as follows~\cite{Hou_naka_order}:
\begin{equation}\label{channel ordering}
{\left| {{{\tilde h}_1}} \right|^2} < {\left| {{{\tilde h}_2}} \right|^2} <  \cdots  < {\left| {{{\tilde h}_{{W}}}} \right|^2}.
\end{equation}
We then turn our attention to the RIS design. It is assumed that the RISs mainly focus on providing maximum channel gain to the prioritized user for enhancing the SE. Without loss of generality, we assume that the prioritized user is the one having the best ordered channel gain.

In this article, in order to simultaneously control multiple RISs, the global CSI is assumed to be perfectly available at the RIS controller.
Since user $W$ is the prioritized user, we aim for maximizing users' received power by designing the phase shifts and reflection amplitude coefficients of RISs as follows:
\begin{equation}\label{define of the passive beamforming}
\begin{aligned}
& Max({{\bf{g}}_{W}}{{\bf{\Phi }}}{{\bf{h}}} + r_{W})\\
& subject\;to\;{\beta _{1}} \cdots {\beta _{N}} = 1\\
& {\theta _{1}} \cdots {\theta _{N}} \in \left[ {0,2\pi } \right).
\end{aligned}
\end{equation}
Thus by utilizing our signal alignment technique, our objective can be achieved by phase-shifting the signals received at the RISs, which is capable of significantly improving the received power.

Thus, we first define a channel vector as follows:
\begin{equation}\label{channel matrix }
{\bf{\tilde h}} = \left[ {\begin{array}{*{20}{c}}
{{g_{W,1}}{h_{1}}{\phi _1}}& \cdots &{{g_{W,N}}{h_{N}}{\phi _N}}
\end{array}} \right].
\end{equation}
Hence, the design of the $n$-th RIS can be expressed as
\begin{equation}\label{RIS design first}
{\Theta}({{{\bf{\tilde h}}}_n}) = \Theta ({r_W}),
\end{equation}
where $\Theta ( \cdot ) = \arg ( \cdot )$ denotes the angle of the element, and ${{{\bf{\tilde h}}}_n}$ denotes the $n$-th element of ${{{\bf{\tilde h}}}}$.

We then generate the effective vector of user $W$ as follows:
\begin{equation}\label{channel matrix effective before phase test}
{\bf{\bar h}} = \left[ {\begin{array}{*{20}{c}}
{{g_{W,1}}{h_{1}}}& \cdots &{{g_{W,N}}{h_{N}}}
\end{array}} \right].
\end{equation}
Thus, the phase shifts of the RISs can be further transformed into
\begin{equation}\label{RIS phase shift design}
{\bf{\Phi }}{\rm{ = }}\Theta {({\bf{\bar h}})^{ - 1}}\Theta ({r_W}).
\end{equation}

Since the phase shifts are designed for the prioritized user $W$, the effective channel gain for user $v$ can be written as ${\left| {{{\bar h}_v}} \right|^2} = {\left( {\sum\limits_{n = 1}^N {} {g_{v,n}}{ \theta_n}{h_n} } d_1^{ - \frac{{{\alpha _l}}}{2}}d_{2,v}^{ - \frac{{{\alpha _l}}}{2}}  +{r_v} d_{3,v}^{ - \frac{{{\alpha _n}}}{2}} \right)^2}$, which cannot be evaluated. However, for the effective channel gain we have:
\begin{equation}\label{range of poor user}
0 \le {\left| {{{\bar h}_v}} \right|^2} \le {\left| {{{\tilde h}_W}} \right|^2}.
\end{equation}
We then consider the situation that two users, i.e. user $W$ and user $v$ having indexes of $v < W$, are paired to perform NOMA. 

\subsection{New Channel Statistics}
In this subsection, we derive new channel statistics for the proposed RIS-aided NOMA network, which will be used for evaluating the OPs and ergodic rates in the following subsections.
\begin{lemma}\label{lemma1:new state of effective channel gain}
Let us assume that the fading parameters of the elements in ${{\rm \bf{h}}}$ and ${{\rm \bf{g}}_{W}}$ are $m_1$ and $m_W$, respectively. The elements of the channel vectors are independently and identically distributed (i.i.d.). On the one hand, the worst-case distribution of the effective channel gain of user $W$ can be formulated as
\begin{equation}\label{New Gamma distribution in Lemma_w-th user}
\left| {{{{\rm{\tilde h}}}_{W,l}}} \right|^2 \sim \Gamma \left(  {\frac{{{{\left( {N{{\left( {{d_1}{d_{2,W}}} \right)}^{ - {\alpha _l}}} + d_{3,W}^{ - {\alpha _n}}} \right)}^2}}}{{N{m_l}{{\left( {{d_1}{d_{2,W}}} \right)}^{ - 2{\alpha _l}}} + d_{3,W}^{ - 2{\alpha _n}}}},\frac{{N{m_l}{{\left( {{d_1}{d_{2,W}}} \right)}^{ - 2{\alpha _l}}} + d_{3,W}^{ - 2{\alpha _n}}}}{{N{{\left( {{d_1}{d_{2,W}}} \right)}^{ - {\alpha _l}}} + d_{3,W}^{ - {\alpha _n}}}}}    \right),
\end{equation}
where $\Gamma \left( \cdot, \cdot \right)$ represents the Gamma distribution, and ${m_l} = \frac{{\left( {1 + {m_1} + {m_W}} \right)}}{{{m_1}{m_W}}}$. On the other hand, the best-case distribution of the effective channel gain of user $W$ can be expressed by
\begin{equation}\label{New Gamma distribution in Lemma_w-th user upper bound}
\left| {{{{\rm{\tilde h}}}_{W,u}}} \right|^2 \sim \Gamma \left(  {\frac{{{{\left( {{N^2}{{\left( {{d_1}{d_{2,W}}} \right)}^{ - {\alpha _l}}} + d_{3,W}^{ - {\alpha _n}}} \right)}^2}}}{{{N^2}{m_u}{{\left( {{d_1}{d_{2,W}}} \right)}^{ - 2{\alpha _l}}} + d_{3,W}^{ - 2{\alpha _n}}}},\frac{{{N^2}{m_u}{{\left( {{d_1}{d_{2,W}}} \right)}^{ - 2{\alpha _l}}} + d_{3,W}^{ - 2{\alpha _n}}}}{{{N^2}{{\left( {{d_1}{d_{2,W}}} \right)}^{ - {\alpha _l}}} + d_{3,W}^{ - {\alpha _n}}}}}   \right),
\end{equation}
where $m_u=\frac{1+Nm_1+Nm_W}{{m_1} {m_W}}$
\begin{proof}
Please refer to Appendix A.
\end{proof}
\end{lemma}

\subsection{Outage Probability}

In this article, the OP of user $W$ is defined by
\begin{equation}\label{Outage Defination}
\begin{aligned}
& {P_w} = \mathbb{P} \left( {{{\log }_2}(1 + SIN{R_{W \to v}}) < {R_v}} \right) +\\
& \mathbb{P} \left( {{{\log }_2}(1 + SIN{R_{W \to v}}) > {R_v},{{\log }_2}(1 + SIN{R_W}) < {R_W}} \right),
\end{aligned}
\end{equation}
where ${{R_W}}$ and $R_v$ represent the target rates of user $W$ and user $v$, respectively.

We then focus our attention on the SINR analysis of user $W$ having the best channel gain. The cell-centre user $W$ first decodes the signal of the cell-edge user $v$ with the following SINR:
\begin{equation}\label{SINR v decode w}
SIN{R_{W \to v}} = \frac{{{{\left| {{{\tilde h}_W}} \right|}^2}{p}\alpha _v^2}}{{{{\left| {{{\tilde h}_W}} \right|}^2}{p}\alpha _W^2 + {\sigma ^2}}}.
\end{equation}
Once the signal of user $v$ is decoded successfully, user $W$ decodes its own signal at an SINR of:
\begin{equation}\label{SINR_v}
SIN{R_W} = \frac{{{{\left| {{{\tilde h}_W}} \right|}^2}{p}\alpha _W^2}}{{{\sigma ^2}}}.
\end{equation}

Let us now turn our attention to calculating the OP of user $W$ based on Theorems $1$ and $2$.
\begin{theorem}\label{Theorem1:Outage W user closed form by incomlete gamma}
\emph{Assuming that $ \alpha _v^2 - \alpha _W^2{\varepsilon _v} > 0$, the worst-case of the closed-form OP expression of user $W$ can be expressed as}
\begin{equation}\label{outage analytical results W in theorem1}
\begin{aligned}
{P_{W,u}} = \frac{\gamma {\left( {{k_1},\frac{{{I_{w*}}}}{{{ \lambda_1}}}} \right)^{W}}}{{\Gamma {{({k_1})}}^W}} ,
\end{aligned}
\end{equation}
\emph{where ${I_{W*}} = \max \left\{ {{I_w},{I_v}} \right\}$, ${I_v} = \frac{{{\varepsilon _v}{\sigma ^2}}}{{{p}\left( {\alpha _v^2 - \alpha _W^2{\varepsilon _v}} \right)}}$, ${I_W} = \frac{{{\varepsilon _W}{\sigma ^2}}}{{{p}\alpha _W^2}}$, ${\varepsilon _v} = {2^{{R_v}}} - 1$, ${\varepsilon _W} = {2^{{R_W}}} - 1$, $ {k_1} = \frac{{{{\left( {N{{\left( {{d_1}{d_{2,W}}} \right)}^{ - {\alpha _l}}} + d_{3,W}^{ - {\alpha _n}}} \right)}^2}}}{{N{m_l}{{\left( {{d_1}{d_{2,W}}} \right)}^{ - 2{\alpha _l}}} + d_{3,W}^{ - 2{\alpha _n}}}} $, ${\lambda _1} = \frac{{N{m_l}{{\left( {{d_1}{d_{2,W}}} \right)}^{ - 2{\alpha _l}}} + d_{3,W}^{ - 2{\alpha _n}}}}{{N{{\left( {{d_1}{d_{2,W}}} \right)}^{ - {\alpha _l}}} + d_{3,W}^{ - {\alpha _n}}}}$, and $\gamma (,)$ represents the lower incomplete Gamma function.}
\begin{proof}
Please refer to Appendix B.
\end{proof}
\end{theorem}

\begin{theorem}\label{Theorem2:Outage W user closed form by incomlete gamma lower bound}
\emph{Assuming that $ \alpha _v^2 - \alpha _W^2{\varepsilon _v} > 0$, the best-case of the closed-form OP expression of user $W$ can be expressed as}
\begin{equation}\label{outage analytical results W in theorem1 lower bound}
\begin{aligned}
{P_{W,l}} = \frac{\gamma {\left( {{k_2},\frac{{{I_{w*}}}}{{{ \lambda_2}}}} \right)^{W}}}{{\Gamma {{({k_2})}}^W}} ,
\end{aligned}
\end{equation}
\emph{where $ {k_2} = \frac{{{{\left( {{N^2}{{\left( {{d_1}{d_{2,W}}} \right)}^{ - {\alpha _l}}} + d_{3,W}^{ - {\alpha _n}}} \right)}^2}}}{{{N^2}{m_u}{{\left( {{d_1}{d_{2,W}}} \right)}^{ - 2{\alpha _l}}} + d_{3,W}^{ - 2{\alpha _n}}}} $, and ${\lambda _2} = \frac{{{N^2}{m_u}{{\left( {{d_1}{d_{2,W}}} \right)}^{ - 2{\alpha _l}}} + d_{3,W}^{ - 2{\alpha _n}}}}{{{N^2}{{\left( {{d_1}{d_{2,W}}} \right)}^{ - {\alpha _l}}} + d_{3,W}^{ - {\alpha _n}}}}$.}
\begin{proof}
Similar to Appendix B, Theorem 2 can be readily proved.
\end{proof}
\end{theorem}

It is however quite challenging to directly obtain engineering insights from~\eqref{outage analytical results W in theorem1} and~\eqref{outage analytical results W in theorem1 lower bound} due to the $W$-th power of the lower incomplete Gamma function. Thus, in order to gain further insights in the high-SNR regime, the approximate behaviors are analyzed, when the SNR is sufficiently high, i.e. when the transmit SNR obeys ${\frac{p}{\sigma^2}  \to \infty }$.
\begin{corollary}\label{corollary1:Outage v user asymptotic}
\emph{Assuming that $ \alpha _v^2 - \alpha _W^2{\varepsilon _v} > 0$, the worst-case and best-case of the OP can be approximated in closed form by}
\begin{equation}\label{asymptotic result v user in corollary1}
\begin{aligned}
{{\bar P}_{W,u}} & = {\left( {\sum\limits_{s = 0}^\infty  {\frac{1}{{\Gamma \left( {{k_1} + s + 1} \right)}}} {{\left( {\frac{{{I_{W*}}}}{{{\lambda _1}}}} \right)}^s}} \right)^W} \sum\limits_{j = 0}^W { {\begin{pmatrix}
W\\
j
\end{pmatrix}} {{( - 1)}^j}} {\left( {\frac{{{I_{W*}}}}{{{\lambda _1}}}} \right)^{{k_1}W + j}},
\end{aligned}
\end{equation}
and
\begin{equation}\label{asymptotic result v user in corollary1 lower bound}
\begin{aligned}
{{\bar P}_{W,l}} & = {\left( {\sum\limits_{s = 0}^\infty  {\frac{1}{{\Gamma \left( {{k_2} + s + 1} \right)}}} {{\left( {\frac{{{I_{W*}}}}{{{\lambda _2}}}} \right)}^s}} \right)^W} \sum\limits_{j = 0}^W { {\begin{pmatrix}
W\\
j
\end{pmatrix}} {{( - 1)}^j}} {\left( {\frac{{{I_{W*}}}}{{{\lambda _2}}}} \right)^{{k_2}W + j}}.
\end{aligned}
\end{equation}
\begin{proof}
Please refer to Appendix C.
\end{proof}
\end{corollary}

The diversity orders of the prioritized user $W$ can be obtained for evaluating the slope of the OP in the following Propositions.
\begin{proposition}\label{proposition1: v diversity order}
\emph{Based on \textbf{Corollary~\ref{corollary1:Outage v user asymptotic}}, the diversity orders can be determined by using the approximate results, explicitly the worst-case and best-case on the diversity order of user $W$ supported by the proposed RIS-aided NOMA network are given by}
\begin{equation}\label{diversity order of w}
{d_{W,u}} =  - \mathop {\lim }\limits_{\frac{{{p}}}{{{\sigma ^2}}} \to \infty } \frac{{\log {{\bar P}_{W,u}}}}{{\log \frac{{{p}}}{{{\sigma ^2}}}}} \approx {k_1}W ,
\end{equation}
and
\begin{equation}\label{diversity order of w lower bound}
{d_{W,l}} =  - \mathop {\lim }\limits_{\frac{{{p}}}{{{\sigma ^2}}} \to \infty } \frac{{\log {{\bar P}_{W,l}}}}{{\log \frac{{{p}}}{{{\sigma ^2}}}}} \approx {k_2}W .
\end{equation}
\end{proposition}

\begin{remark}\label{remark1:impact of N on diversity order}
The results of~\eqref{diversity order of w} demonstrate that the diversity orders can be approximated by $\frac{{NW}}{{{m_l}}}$ for the prioritized user $W$, when the number of RISs is high enough. It is also demonstrated that increasing the number of RISs and carefully pairing the NOMA users is capable of significantly improving the outage performance.
\end{remark}

\begin{remark}\label{remark2:impact of fading environment on diversity order}
Assuming that ${m_1} \to \infty $, which indicates a strong LoS link between the BS as well as the RISs, and provided that the number of RISs is high enough, the diversity orders on both the best-case and worst-case of the prioritized user $W$ can be approximated by $NW{m_W}$.
\end{remark}

\begin{remark}\label{remark3:impact of the number of users on diversity order}
Since the users are ordered by their effective channel gain, and based on the results of~\eqref{diversity order of w}, in order to minimize the OP of the paired NOMA users, it is preferable to pair the users having the best and the second best effective channel gains.
\end{remark}

\begin{remark}\label{remark4:lower bound}
Assuming that the number of RISs is high enough, and based on the results of~\eqref{diversity order of w lower bound}, the diversity order of the worst-case on the OP can be further approximated by $\frac{NW{m_1}{m_W}}{{m_1}+{m_W}}$. Again, assuming that ${m_1} \to \infty $, both the worst-case and best-case on the diversity order of user $W$ can be approximated by $NW{m_W}$, which indicates that the diversity order of both the best-case and worst-case of the OP are identical.
\end{remark}

\begin{remark}\label{remark5:direct link domain}
Assuming that the BS-user link of user $W$ is the dominant component, where the path loss exponent $\alpha_N=\alpha_L$ as well as ${d_1}{d_{2,W}}>>{d_{3,W}}$, the diversity order of both the best-case and worst-case are $W$.
\end{remark}

Due to the impact of hostile fading environments in NG networks, it is worth mentioning that no BS-user link may be available between the BS and the paired NOMA users, and the approximate result mainly depends on the $0$-th ordered element in~\eqref{asymptotic result v user in corollary1} and~\eqref{asymptotic result v user in corollary1 lower bound}. Thus, we continue by providing basic numerical insights using the following Corollary.
\begin{corollary}\label{corollary3:Outage w and v user asymptotic in deep fading}
\emph{Due to the hostile fading environment between the BS and the users in NG networks, and assuming that $ \alpha _v^2 - \alpha _W^2{\varepsilon _v} > 0$, the $0$-th ordered elements in terms of the worst-case and best-case on the approximate OP of user $W$ are given by}
\begin{equation}\label{asymptotic result v user in corollary3 deep fading}
\begin{aligned}
{{\tilde P}_{W,u}} = {\frac{1}{{\Gamma \left( {{\varphi_1}  + 1} \right)^W}}} \sum\limits_{j = 0}^{W} { {\begin{pmatrix}
{W}\\
j
\end{pmatrix}} {{( - 1)}^j}} {\left( {\frac{{{I_{W*}}}}{{{m_l}{{\left( {{d_1}{d_{2,W}}} \right)}^{ - {\alpha _l}}}}}} \right)^{ - ({\varphi _1}W + j)}},
\end{aligned}
\end{equation}
and
\begin{equation}\label{asymptotic result v user in corollary3 deep fading lower bound}
\begin{aligned}
{{\tilde P}_{W,l}} = {\frac{1}{{\Gamma \left( {{\varphi_2}  + 1} \right)^W}}} \sum\limits_{j = 0}^{W} { {\begin{pmatrix}
{W}\\
j
\end{pmatrix}} {{( - 1)}^j}} {\left( {\frac{{{I_{W*}}}}{{{m_u}{{\left( {{d_1}{d_{2,W}}} \right)}^{ - {\alpha _l}}}}}} \right)^{ - ({\varphi _2}W + j)}},
\end{aligned}
\end{equation}
\emph{where $\varphi_1  = \frac{{N}}{{{m_l}}}$, and $\varphi_2  = \frac{{N^2}}{{{m_u}}}$.}
\end{corollary}

\begin{remark}\label{remark4:insights of deep fading environment}
Assuming that no BS-user links are expected between the BS and the prioritized NOMA user, based on results of~\eqref{asymptotic result v user in corollary3 deep fading}, the best-case and worst-case on the diversity orders of user $W$ are seen to be $\frac{{N^2 W}}{{{m_u}}}$ and $\frac{{NW}}{{{m_l}}}$, respectively.
\end{remark}

\subsection{Ergodic Rate}
The ergodic rate is a salient performance metric related to the SE and EE. Therefore, we focus our attention on analyzing the ergodic rate of user $W$. The approximate ergodic rate expressions of user $W$ are given in the following Theorems.

\begin{theorem}\label{theorem3:ergodic rate W-th user}
\emph{Assuming that $N$ RISs simultaneously serve user $W$, and $ \alpha _v^2 - \alpha _W^2{\varepsilon _v} > 0$, the worst-case on the ergodic rate of user $W$ can be expressed in closed form as follows:}
\begin{equation}\label{asympto W-th erogodic rate in corollary4}
\begin{aligned}
& {R_{W,l}}  = \sum\limits_{s = 0}^W { \begin{pmatrix}
W\\
s
\end{pmatrix} } {( - 1)^s}\frac{1}{{\bar k}} \sum\limits_{{a_1} +  \ldots  + {a_{\bar k}} = s} {} \begin{pmatrix}
s\\
{a_1}, \ldots ,{a_{\bar k}}
\end{pmatrix} {\prod\limits_{t = 1}^{\bar k} {\left( {\frac{{{{\left( C \right)}^{t - 1}}}}{{(t - 1)!}}} \right)} ^{{a_t}}}  \\
& \times \left( {\exp (Cs)Ei( - Cs) + \sum\limits_{i = 1}^{\left( {t - 1} \right){a_t}} {{{( - 1)}^{i - 1}}(i - 1)!{{(Cs)}^i}} } \right),
\end{aligned}
\end{equation}
where $C = \frac{{{\sigma ^2}}}{{{\lambda _1}{p}\alpha _W^2}}$, and $\bar k =\left[ {{k_1}} \right]$ is obtained by rounding ${{k_1}}$ to the nearest integer.
\begin{proof}
Please refer to Appendix D.
\end{proof}
\end{theorem}

Similarly, the best-case on the ergodic rate of user $W$ is formulated in the following Theorem.
\begin{theorem}\label{theorem3:ergodic rate W-th user upper bound}
\emph{Assuming that $N$ RISs simultaneously serve user $W$, and $ \alpha _v^2 - \alpha _W^2{\varepsilon _v} > 0$, the best-case on the ergodic rate of user $W$ can be expressed in closed form as follows:}
\begin{equation}\label{asympto W-th erogodic rate in corollary4 upper bound}
\begin{aligned}
& {R_{W,u}}  = \sum\limits_{s = 0}^W { \begin{pmatrix}
W\\
s
\end{pmatrix} } {( - 1)^s}\frac{1}{{\bar k}} \sum\limits_{{a_1} +  \ldots  + {a_{\bar k_u}} = s} {} \begin{pmatrix}
s\\
{a_1}, \ldots ,{a_{\bar k_u}}
\end{pmatrix} {\prod\limits_{t = 1}^{\bar k} {\left( {\frac{{{{\left( C_u \right)}^{t - 1}}}}{{(t - 1)!}}} \right)} ^{{a_t}}}  \\
& \times \left( {\exp (C_u s)Ei( - C_u s) + \sum\limits_{i = 1}^{\left( {t - 1} \right){a_t}} {{{( - 1)}^{i - 1}}(i - 1)!{{(C_u s)}^i}} } \right),
\end{aligned}
\end{equation}
where $C_u = \frac{{{\sigma ^2}}}{{{\lambda _2}{p}\alpha _W^2}}$, and $\bar k_u = \left[ {{k_2}} \right]$.
\begin{proof}
Similar to Appendix D, the results in~\eqref{asympto W-th erogodic rate in corollary4 upper bound} can be readily obtained.
\end{proof}
\end{theorem}

To gain deep insights into the system's performance, the high-SNR slope, as the key parameter determining the ergodic rate in the high-SNR regime, is worth estimating. Therefore, we first express the high-SNR slope as
\begin{equation}\label{High SNR slope of m}
S_\infty ^{W} = -\mathop {\lim }\limits_{ \frac{p}{\sigma^2} \to \infty } \frac{{{R_{W}}}}{{{{\log }_2}\left( {1 + \frac{p}{\sigma^2}} \right)}}.
\end{equation}

\begin{proposition}\label{proposition4: m high SNR slopes in massive LIS}
\emph{By substituting~\eqref{asympto W-th erogodic rate in corollary4} and~\eqref{asympto W-th erogodic rate in corollary4 upper bound} into~\eqref{High SNR slope of m}, the high-SNR slope of user $W$ is given by}
\begin{equation}\label{high SNR slope in massive LIS}
S_\infty ^{W} = 1.
\end{equation}
\end{proposition}

\begin{remark}\label{remark6:impact of ergodic rate in massive LIS}
The results of~\eqref{high SNR slope in massive LIS} illustrate that the slope of the ergodic rate in the proposed RIS-aided NOMA network is one, which is not affected by the number of RISs.
\end{remark}

Based on the passive beamforming weight design at the RISs, the distribution of NOMA user $v$, having the lower received power, cannot be evaluated. Hence, we only provide the associated SINR analysis for simplicity. By relying on the NOMA protocols, user $v$ treats the signal from user $W$ as interference, and the SINR is given by
\begin{equation}\label{SINR_poor user}
SIN{R_v} = \frac{{({{\left| {{{\bf{g}}_v}{\bf{\Phi h}} }+ {r_v} \right|}^2}) {\alpha _v^2}{p}}}{{({{\left| {{{\bf{g}}_v}{\bf{\Phi h}} }+{r_v} \right|}^2}) \alpha _W^2{p} + {\sigma ^2}}}.
\end{equation}
Since the elements in ${\bf{\Phi}}$ are gleaned from random variables, and based on the insights from~\cite{Yi_anlog_beam}, the SINR of user $v$ can be further approximated as:
\begin{equation}\label{SINR_poor user_approx}
SIN{R_v} = \frac{{({{\left| {{{\bf{g}}_v}{\bf{h}}} \right|}^2}{G_N}(\bar \theta ) + {{\left| {{r_v}} \right|}^2})\alpha _v^2p}}{{({{\left| {{{\bf{g}}_v}{\bf{h}}} \right|}^2}{G_N}(\bar \theta ) + {{\left| {{r_v}} \right|}^2})\alpha _W^2p + {\sigma ^2}}},
\end{equation}
where ${{G_N}(\bar \theta )}$ denotes the normalized \emph{Fej{\`e}r Kernel} function with parameter $N$. Note that ${{G_N}(\bar \theta )}$ has a period of two, hence $\bar \theta $ is uniformly distributed over $\left[ { - 1,1} \right]$. Thus, the ergodic rate of user $v$ can be expressed as follows.
\begin{theorem}\label{theorem4:ergodic rate v-th user both}
\emph{Assuming that $N$ RISs simultaneously serve user $W$, and $ \alpha _v^2 - \alpha _W^2{\varepsilon _v} > 0$, the worst-case and best-case on the ergodic rate of user $v$ can be expressed as follows:}
\begin{equation}\label{asympto v-th erogodic rate in theo lower bound}
\begin{aligned}
{R_{v,l}} = \frac{1}{{\ln \left( 2 \right)}}\int_0^{\frac{{\alpha _v^2}}{{\alpha _W^2}}} {} \frac{{1 - {F_{v,l}}\left( x \right)}}{{1 + x}}dx,
\end{aligned}
\end{equation}
and
\begin{equation}\label{asympto v-th erogodic rate in theo upper bound}
\begin{aligned}
{R_{v,u}} = \frac{1}{{\ln \left( 2 \right)}}\int_0^{\frac{{\alpha _v^2}}{{\alpha _W^2}}} {} \frac{{1 - {F_{v,u}}\left( x \right)}}{{1 + x}}dx,
\end{aligned}
\end{equation}
where ${F_{v,l}}\left( x \right) = {\left( {\frac{{\gamma \left( {{k_1},{C_{v,l}}} \right)}}{{\Gamma ({k_1})}}} \right)^v}$, ${F_{v,u}}\left( x \right) = {\left( {\frac{{\gamma \left( {{k_2},{C_{v,u}}} \right)}}{{\Gamma ({k_2})}}} \right)^v}$, ${C_{v,l}} = \frac{{{\sigma ^2}x}}{{{\lambda _1}{G_N}(\bar \theta )p\left( {\alpha _v^2 - \alpha _W^2x} \right)}}$, and ${C_{v,u}} = \frac{{{\sigma ^2}x}}{{{\lambda _2}{G_N}(\bar \theta )p\left( {\alpha _v^2 - \alpha _W^2x} \right)}}$.
\begin{proof}
Similar to Appendix D, the results can be readily derived.
\end{proof}
\end{theorem}

\begin{remark}\label{remark6:impact of ergodic rate in massive LIS}
Let us assume that $\bar \theta \to 0$, indicating that the paired NOMA users share an identical channel vector, the Fej{\`e}r Kernel function can be considered as one. Hence, based on the insights in~\cite{Hou_Single_UAV}, the best-case and worst-case on the ergodic rate of user $v$ may approach $R_v={\rm{log}}_2\left(1+\frac{\alpha_v^2}{\alpha_W^2}\right)$ in the high-SNR regime.
\end{remark}

In order to provide further insights for RIS-aided NOMA networks, the ergodic rate of the paired users is also analysed in the OMA scenario using TDMA. The OMA benchmark adopted in this article relies on supporting user $W$ and user $v$ in a pair of identical time slots. In each time slot, the RISs provide access only for one of the users. Thus, the channel capacity of user $W$ in the OMA scenario can be expressed as
\begin{equation}\label{OMA benchmark for user W}
{R_{W,O}} = \mathbb{E} \left\{ {\frac{1}{2}{{\log }_2}\left( {1 + SN{R_{W,O}}} \right)} \right\} ,
\end{equation}
where $SIN{R_{W,O}} = \frac{{\left( {{{\left( {\sum\limits_{n = 1}^N {} \left| {{{{g}}_{W,n}}{{{h}}_n}} \right|} \right)}^2}{\left( {{d_1}{d_{2,W}}} \right)^{ - {\alpha _l}}} + {{\left| {{r_W}} \right|}^2} {d_{3,W}^{ - {\alpha _n}}}  } \right)p}}{{{\sigma ^2}}}$.
Similarly, the channel capacity of user $v$ can be expressed as
\begin{equation}\label{OMA benchmark for user v}
{R_{v,O}} = \mathbb{E} \left\{ \frac{1}{2} {{{\log }_2}\left( {1 + SN{R_{v,O}}} \right)} \right\} ,
\end{equation}
where $SIN{R_{v,O}} = \frac{{\left( {{{\left( {\sum\limits_{n = 1}^N {} \left| {{{{g}}_{v,n}}{{{h}}_n}} \right|} \right)}^2}{\left( {{d_1}{d_{2,v}}} \right)^{ - {\alpha _l}}}  + {{\left| {{r_v}} \right|}^2}{d_{3,v}^{ - {\alpha _n}}} } \right)p}}{{{\sigma ^2}}}$.

\subsection{Spectrum Efficiency and Energy Efficiency}

Based on the analysis of the previous two subsections, a tractable SE expression can be formulated in the following Proposition.
\begin{proposition}\label{proposition5: spectrum efficiency}
\emph{The SE of the proposed RIS-aided NOMA network is given by }
\begin{equation}\label{spectrum efficiency}
S = {{R_{v}}}+  {{R_{W}}} .
\end{equation}
\end{proposition}

In NG networks, EE is an important performance metric. Thus, based on insights gleaned from~\cite{EE_model_massive_MIMO}, we first model the total power dissipation of the proposed RIS-aided NOMA network as
\begin{equation}\label{total energy massive LIS}
{P_{e}} = {P_{{\rm{B,s}}}} + 2 P_{\rm{U}} + p{{{\varepsilon _b}}} + {N}{P_L},
\end{equation}
where ${P_{{\rm{B,s}}}}$ is the static hardware power consumption of the BS, ${{\varepsilon _b}}$ denotes the efficiency of the power amplifier at the BS, $P_{\rm{U}}$ is the power consumption of each user, and ${P_L}$ represents the power consumption of each RIS controller. Thus, the EE of the proposed network is given by the following Proposition.

\begin{proposition}\label{proposition7: energy efficiency massive LIS}
\emph{The EE of the proposed RIS-aided NOMA network is }
\begin{equation}\label{energy efficiency massive LIS}
{\Theta _{EE}} = \frac{S}{{{P_{e}}}},
\end{equation}
where $S$ and $P_{e}$ are obtained from~\eqref{spectrum efficiency} and~\eqref{total energy massive LIS}, respectively.
\end{proposition}

\section{Numerical Studies}

In this section, numerical results are provided for the performance evaluation of the proposed network.
Monte Carlo simulations are conducted for verifying the accuracy of our analytical results. The bandwidth of the DL is set to $BW=1$ MHz, and the power of the AWGN is set to $\sigma^2= −-174+ 10 {\rm log}_{10}(BW)$ dBm. The power attenuation at the reference distance is set to~-30~dB, and the reference distance is set to 1 meter.
Note that the LoS and NLoS links are indicated by the Nakagami fading parameter, where $m = 1$ and $m >1$ are for NLoS and for LoS links, respectively. The target rates are $R_W=1.5$ and $R_v=1$ bits per channel use (BPCU).
The power allocation factors of the paired NOMA users are set to $\alpha_v^2=0.6$ and $\alpha_W^2=0.4$. The number of users is set to $W=2$, and $v=1$. The fading environments are set to $m_1=m_W=m_v=1$. The length of the BS-RIS link is set to $d_1=60$m. The length of the RIS-user links are set to $d_{2,W}=80$m and $d_{2,v}=100$m, and these of the BS-user links are set to $d_{3,W}= d_{3,v}=100$m. The path loss exponents of the reflected BS-RIS-user and the direct BS-user links are set to $\alpha_l=2.2$ as well as $\alpha_n=3.5$, respectively, unless otherwise stated.

\begin{figure}[t!]
\centering
\includegraphics[width =3in]{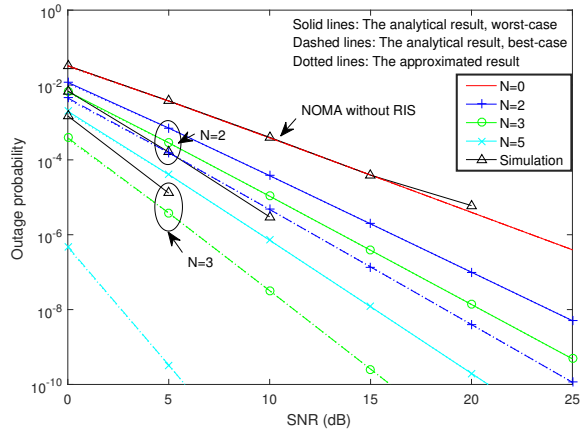}
\caption{OP of the RIS-aided NOMA network versus the SNR parameterized by the number of RISs. The analytical and approximate results are calculated from~\eqref{outage analytical results W in theorem1},~\eqref{outage analytical results W in theorem1 lower bound},~\eqref{asymptotic result v user in corollary1} as well as~\eqref{asymptotic result v user in corollary1 lower bound}, respectively. }
\label{SISO_number of IRS_fig1}
\vspace{-0.2in}
\end{figure}

\emph{1) Impact of the Number of RISs:} In Fig.~\ref{SISO_number of IRS_fig1}, we focus our attention on the OP of the RIS-aided NOMA network. The solid curves and dashed curves represent the worst-case and best-case of the analytical results, respectively.
We can see that as the number of RISs serving user $W$ increases, the OP decreases. This is due to the fact that, as more RISs are employed, the received signal power can be significantly increased as a benefit of the increased diversity order.
Observe that the slope of the curves increases with the number of RISs, which validates our~\textbf{Remark~\ref{remark1:impact of N on diversity order}}.
Let us assume that $d_{2,W}=d_{3,W}$ and $\alpha_N=\alpha_L$, then the minimum diversity order that can be obtained is $1$ for the case of $m_1=m_W=1$ and $N=1$, which is identical to that of the non-RIS-aided networks. Observe that as expected the simulation results are located between the best and worst cases, which verifies~\textbf{Remark~\ref{remark4:lower bound}}.

\begin{figure}[t!]
\centering
\includegraphics[width = 3 in]{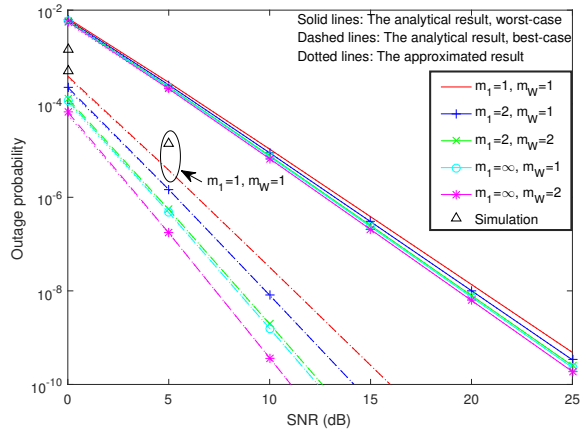}
\caption{OP of the RIS-aided NOMA network versus the SNR parameterized by fading factors. The number of RISs is set to $N=3$.}
\label{SISO_fading_evironments_fig2}
\vspace{-0.2in}
\end{figure}

\emph{2) Impact of Fading Environments:} In Fig.~\ref{SISO_fading_evironments_fig2}, we evaluate the OP of the prioritized user $W$ in different fading environments. As expected, with the transmit power increases, the OP decreases. Observe that both the BS-RIS as well as RIS-user links have an impact on the OP, which is in contrast to the FOD of~\cite{Hou_RIS_MIMO_global_algrithm}, where the fading environment of the RIS-user link has almost no effect on the OP. 


\begin{figure}[t!]
\centering
\includegraphics[width =3 in]{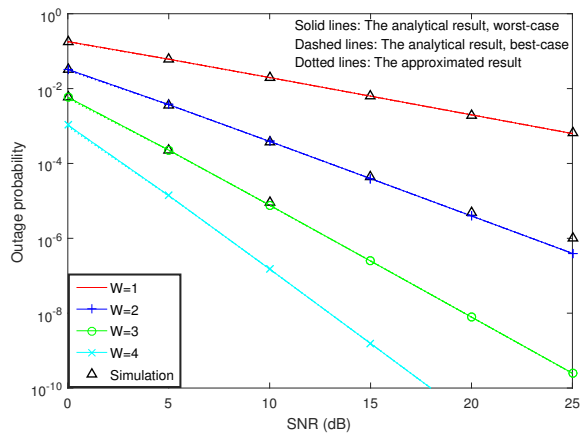}
\caption{OP of the RIS-aided NOMA network versus the SNR parameterized by the number of users. The number of RISs is set to $N=3$. The path loss exponents are set to $\alpha_n=\alpha_l=3$.}
\label{SISO_order_statistic_fig3}
\vspace{-0.2in}
\end{figure}

\emph{3) Impact of the Number of Users:} Let us now study the impact of the number of users in Fig.~\ref{SISO_order_statistic_fig3}. Observe that it is preferable to pair the users having the best effective channel gains for minimizing the OP. 
Based on the results in the high-SNR regime, the diversity order is seen to be significantly enhanced by increasing the number of users, because they experience independent fading channels.
It is also worth noting that the diversity order is $W$, which verified by the insights gleaned from \textbf{Remark~\ref{remark2:impact of fading environment on diversity order}}. This is because when the path loss exponent is $\alpha_l=\alpha_n=3$, the power received from links reflected by the RISs can be nearly ignored.

\begin{figure}[t!]
\centering
\subfigure[Ergodic rate of user $W$, where the analytical results of user $W$ are calculated from~\eqref{asympto W-th erogodic rate in corollary4} as well as~\eqref{asympto W-th erogodic rate in corollary4 upper bound}.]{\label{Ergodic_W user}
\includegraphics[width =3 in]{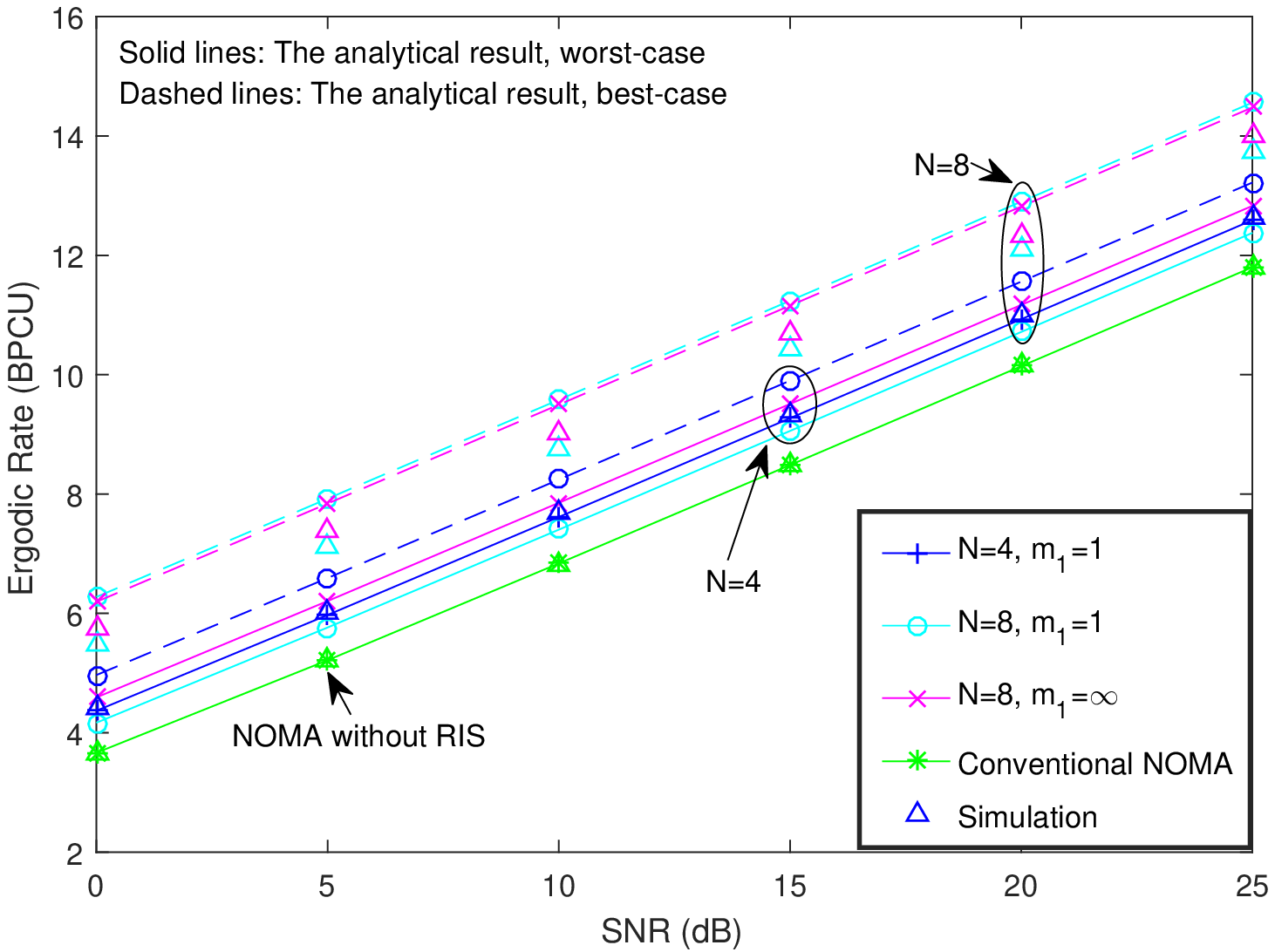}}
\subfigure[Ergodic rate of user $v$, where the channel gains of the best-case and worst-case are derived similar to~\eqref{New Gamma distribution in Lemma_w-th user} and~\eqref{New Gamma distribution in Lemma_w-th user upper bound}.]{\label{Ergodic_v user}
\includegraphics[width =3 in]{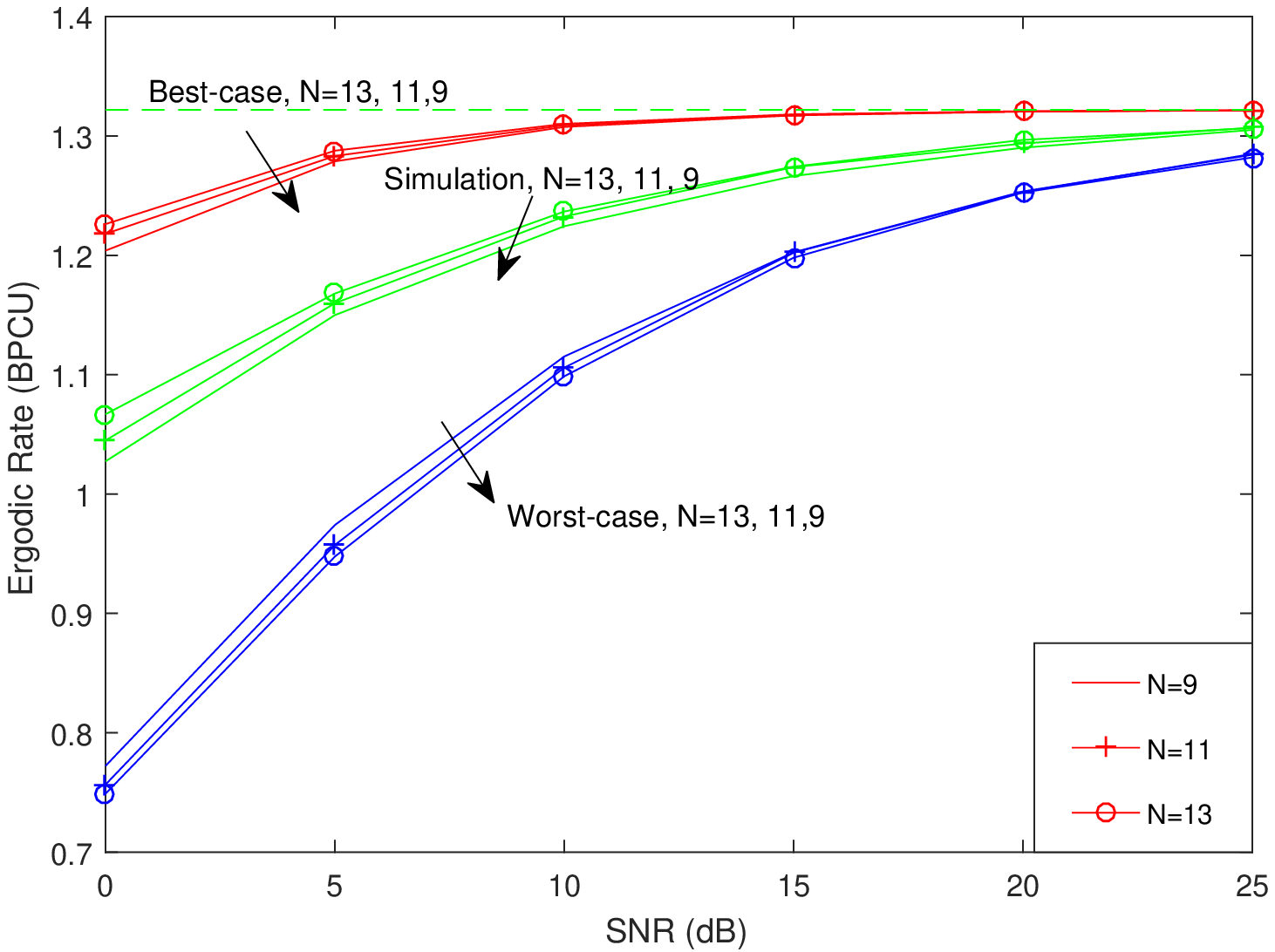}}
\caption{Ergodic rate of paired NOMA users versus transmit SNR.}
\label{SISO_ergodic rate}
\vspace{-0.2in}
\end{figure}

\emph{4) Ergodic Rate:} Fig.~\ref{SISO_ergodic rate} compares the ergodic rates of paired NOMA users versus the SNR parameterized by the fading parameters and by the number of RISs. Several observations can be drawn as follows: 1) Based on the curves in~Fig.~\ref{Ergodic_W user}, we can observe that the LoS links of both the BS-RIS as well as of the RIS-user links increase the ergodic rate of user $W$, where the ergodic rate approaches the best-case for the case of $m_1 \to  \infty$. 2) The triangles are between the best-case and worst-case, which verify the accuracy of our results. 3) As seen from the figure, the high-SNR slope of user $W$ is one, which also verifies~\textbf{Remark~\ref{remark6:impact of ergodic rate in massive LIS}}. 4) The ergodic rate can be significantly increased by employing more RISs, which is because the spatial diversity gain can be significantly increased upon increasing the number of RISs. 5) The ergodic rate of conventional NOMA dispensing with RISs is provided as the benchmark schemes, which can be calculated by setting the number of RISs to $N=0$. 6) Fig.~\ref{Ergodic_v user} evaluates the ergodic rate of the non-prioritized user $v$. Observe that in the high-SNR regime, the slope of user $v$ approaches zero in~Fig.~\ref{Ergodic_v user}, which indicates that the number of RISs has no significant impact on the ergodic rate of user $v$. In TABLE~\ref{DIVERSITY ORDER AND HIGH SNR SLOPE FOR RIS Networks}, we use “D” and “S” to represent the diversity order and high-SNR slope for the case that $N$ is large enough, respectively. It is worth noting that the diversity order of the non-prioritized user $v$ is the optimized result, which can only be obtained by setting $\bar \theta  \to 0$.

\begin{table}
\caption{\\ DIVERSITY ORDER AND HIGH-SNR SLOPE}
\centering
\begin{tabular}{|l|c|c|c|}
\hline
Access Mode & Rx & D & S \\
\hline
\multirow{2}{*}{RIS-aided NOMA}
& $W$ & $\frac{N {m_1} {m_W}W}{m_1 +{m_W}} $ & 1 \\
\cline{2-4}
& $v$ & $\frac{N {m_1} {m_v}v}{m_1 +{m_v}} $ & 0 \\
\hline
\multirow{2}{*}{Conventional NOMA}
& $W$ & $W$ & 1 \\
\cline{2-4}
& $v$ & $v$ & 0 \\
\hline
\multirow{2}{*}{OMA}
& $W$ & $W$ & 0.5 \\
\cline{2-4}
& $v$ & $v$ & 0.5 \\
\hline
\end{tabular}
\label{DIVERSITY ORDER AND HIGH SNR SLOPE FOR RIS Networks}
\end{table}

\begin{figure}[t!]
\centering
\includegraphics[width =3 in]{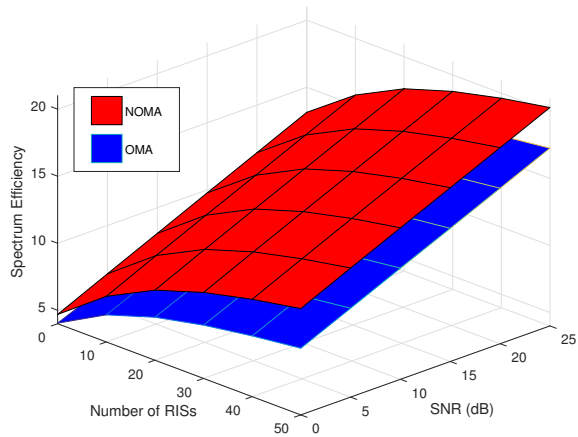}
\caption{SE of both the RIS-aided NOMA and OMA networks versus the SNR and the number of RISs. The fading parameters are set to $m_1=m_W=m_v=3$.}
\label{SISO_compare_OMA}
\vspace{-0.2in}
\end{figure}

\emph{5) Comparing the RIS-aided NOMA to an OMA Network:} In Fig.~\ref{SISO_compare_OMA}, we then evaluate the SE of our RIS-aided NOMA network, as well as that of its OMA counterpart .
The results of the RIS-aided NOMA and OMA networks are derived by $R_{W}+ R_{v}$ and $R_{W,O}+R_{v,O}$, respectively.
We can see that the RIS-aided NOMA network is capable of outperforming its OMA counterpart in terms of its SE by appropriately setting the power allocation factors.
Observe that the SE gap between the RIS-aided NOMA network and its OMA counterpart becomes higher, when the number of RISs is increased, which indicates that it is preferable to employ more RISs for enhancing the SE.

\begin{figure}[t!]
\centering
\includegraphics[width =3 in]{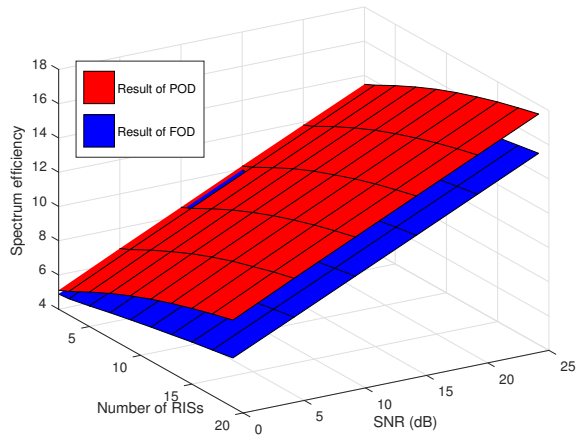}
\caption{SE of the proposed RIS-aided NOMA network versus the SNR and the number of RISs. The results of POD and FOD are calculated from~\eqref{spectrum efficiency} and~\cite{Hou_RIS_MIMO_global_algrithm}. The fading parameters are set to $m_1=m_W=m_v=3$.}
\label{SE fig}
\vspace{-0.2in}
\end{figure}

\emph{6) Comparing the POD to the FOD:} In Fig.~\ref{SE fig}, we evaluate the SE of the proposed POD. The SE of the FOD in~\cite{Hou_RIS_MIMO_global_algrithm} is provided as the benchmark schemes.
Observe from the figure that the SE of the POD is higher than the FOD of~\cite{Hou_RIS_MIMO_global_algrithm}, which indicates that the proposed POD becomes more competitive compared to the FOD. This is due to the fact that the proposed POD is conceived for attainting the maximum network throughput for the prioritized user, which is capable of providing higher SE. By contrast, the FOD is mainly focused on the fairness, which can provide higher throughput for the cell-edge users.

\begin{figure}[t!]
\centering
\includegraphics[width =3in]{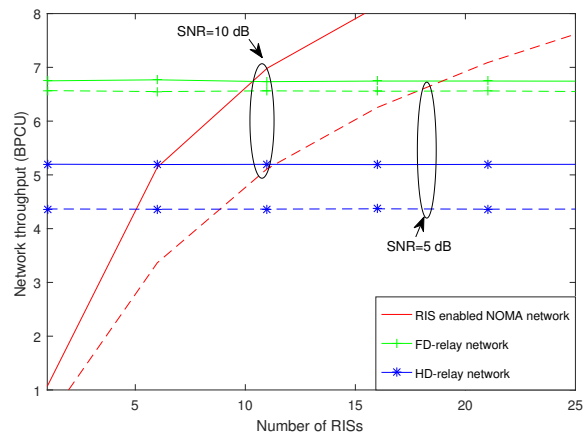}
\caption{Network throughput of the RIS-aided NOMA, HD-relay as well as FD-relay networks versus the number of RISs, where the fading parameters are set to $m_1=m_W=m_v=3$. The loop-back self-interference coefficient of FD-relay is set to ${\epsilon_H}=0.1$. The path loss exponent is set to $\alpha_l=2.5$.}
\label{Compare With AF_DF relay fading fig 6}
\vspace{-0.2in}
\end{figure}

\emph{7) Comparing Half-duplex and Full-duplex Relay Networks:}

In order to provide further engineering insights, combined with the insights inferred from~\cite{yuanwei_cooperative,DF_relaying_outage,cooperative_Yue}, the network throughputs of alternative full-duplex (FD) and half-duplex (HD) cooperative networks are evaluated. We consider the classic relaying protocols, where the transmissions of HD-relaying are divided into two identical phases. By contrast, the FD-relay is suffering from self-interference. It is assumed that both the BS and relay, as well as the users are equipped by a single antenna.
Similar to~\eqref{maximum achievable gain}, we first rank the entire set of $W$ users according to their effective channel gains. We then evaluate the SE of the FD network, where the FD-relay has to decode the signal of the paired NOMA users. Based on the insights from~\textbf{Remark~\ref{remark4:insights of deep fading environment}} for simplicity, it is assumed that no BS-user link exists. Hence, the FD-relay first decodes the signal of user $v$, achieving the following expectation:
\begin{equation}\label{HF decode user v}
{R_{{\rm{F}},v}} = \mathbb{E} \left\{ {{{\log }_2}\left( {1 + \frac{{p{{\left| {{h_{R,1}}} \right|}^2}{d_1^{-\alpha_l}}\alpha _v^2}}{{p{{\left| {{h_{R,1}}} \right|}^2}{d_1^{-\alpha_l}}\alpha _W^2 + {p_d} {\epsilon_H} + {\sigma ^2}}}} \right)} \right\},
\end{equation}
where ${{{\left| {{h_{R,1}}} \right|}^2}}$ denotes the channel gain between the BS and the FD-relay, ${\epsilon_H}$ denotes the self-interference coefficient of the FD-relay itself, and ${{p_d}}$ denotes the transmit power of the relay.
 Then the FD-relay can decode the signal of user $W$ as follows:
\begin{equation}\label{HF decode user W}
{R_{{\rm{F}},W}} = \left\{ {{{\log }_2}\left( {1 + \frac{{p{{\left| {{h_{R,1}}} \right|}^2}{d_1^{-\alpha_l}}\alpha _W^2}}{{{p_d}{\epsilon_H} + {\sigma ^2}}}} \right)} \right\},
\end{equation}

Similarly, assuming that SIC can be also invoked successfully by the paired NOMA users, and thus the non-prioritized user $v$ treats the signal of user $W$ as interference, and the expected data rate can be given by
\begin{equation}\label{FD-user v decode}
{R_v} =\mathbb{E} \left\{ {{{\log }_2}\left( {1 + \frac{{{p_d}{{\left| {{h_{R,v}}} \right|}^2} {d_{2,v}^{-\alpha_l} } \alpha _v^2}}{{{p_d}{{\left| {{h_{R,v}}} \right|}^2}{d_{2,v}^{-\alpha_l} }\alpha _W^2 + {\sigma ^2}}}} \right)} \right\},
\end{equation}
where ${{\left| {{h_{R,v}}} \right|}^2}$ denotes the channel gain between the FD-relay and user $v$. On the other hand, by utilizing SIC technique, the transmission rate of user $W$ is given by
\begin{equation}\label{user W decode user W}
{R_W} = \mathbb{E} \left\{ {{{\log }_2}\left( {1 + \frac{{{p_d}{{\left| {{h_{R,W}}} \right|}^2}{d_{2,W}^{-\alpha_l} }\alpha _W^2}}{{{\sigma ^2}}}} \right)} \right\}.
\end{equation}
More specifically, the size of data rate for user $v$ and user $W$ depend on four kinds of data rates, such as 1) the data rate for the relay to detect user $v$; 2) the data rate for the relay to detect user $W$; 2) The data rate for user $v$; and 3) the data rate for user $W$. Among the FD-relay in the network, based on~\eqref{HF decode user v} to~\eqref{user W decode user W}, the expected rate of the paired NOMA users in the FD-relay network can be given by
\begin{equation}\label{expected rate user v}
{{\bar R}_{{\rm{F}},v}} = \min \left\{ {{R_{{\rm{F}},v}},{R_v}} \right\},
\end{equation}
and
\begin{equation}\label{expected rate user w}
{{\bar R}_{{\rm{F}},W}} = \min \left\{ {{R_{{\rm{F}},W}},{R_W}} \right\}.
\end{equation}
We then consider the HD-relay network, where the expected data rate of the paired NOMA users at the HD-relay can be given by
\begin{equation}\label{HF relay decode v}
{{ R}_{{\rm H},v}} =\mathbb{E} \left\{ {\frac{1}{2}{{\log }_2}\left( {1 + \frac{{p{{\left| {{h_{R,1}}} \right|}^2}{d_1^{-\alpha_l}}\alpha _v^2}}{{p{{\left| {{h_{R,1}}} \right|}^2}{d_1^{-\alpha_l}}\alpha _W^2 + {\sigma ^2}}}} \right)} \right\},
\end{equation}
and
\begin{equation}\label{HF relay decode W}
{{ R}_{{\rm H},W}} =\mathbb{E} \left\{ {\frac{1}{2}{{\log }_2}\left( {1 + \frac{{p{{\left| {{h_{R,1}}} \right|}^2}{d_1^{-\alpha_l}}\alpha _W^2}}{{{\sigma ^2}}}} \right)} \right\}.
\end{equation}
By applying the classic SIC technique, the expected data rate of paired NOMA users can be written as
\begin{equation}\label{HF relay v decode v}
{{\bar R}_v} = \mathbb{E} \left\{ {\frac{1}{2}{{\log }_2}\left( {1 + \frac{{{p_d}{{\left| {{h_{R,v}}} \right|}^2}{d_{2,v}^{-\alpha_l}}\alpha _v^2}}{{{p_d}{{\left| {{h_{R,v}}} \right|}^2}{d_{2,v}^{-\alpha_l}}\alpha _W^2 + {\sigma ^2}}}} \right)} \right\},
\end{equation}
and
\begin{equation}\label{HF relay W decode W}
{{\bar R}_W} =\mathbb{E} \left\{ {\frac{1}{2}{{\log }_2}\left( {1 + \frac{{{p_d}{{\left| {{h_{R,W}}} \right|}^2}{d_{2,W}^{-\alpha_l}}\alpha _W^2}}{{{\sigma ^2}}}} \right)} \right\}.
\end{equation}
Thus, the expected rate of the paired NOMA users in the HD-relay network is given by
\begin{equation}\label{expected rate user v}
{{\bar R}_{{\rm{H}},v}} = \min \left\{ {{ { R}_{{\rm H},v}},{{\bar R}_v}} \right\},
\end{equation}
and
\begin{equation}\label{expected rate user w}
{{\bar R}_{{\rm{H}},W}} = \min \left\{ {{{ R}_{{\rm H},W}},{{\bar R}_W}} \right\}.
\end{equation}

In Fig.~\ref{Compare With AF_DF relay fading fig 6}, we evaluate the network throughput of our RIS-aided NOMA network, as well as of the HD-relay and FD-relay aided networks.
The results of the FD-relay and HD-relay are given by ${\bar R}_{{\rm{F}},v}+{\bar R}_{{\rm{F}},W} $ and ${\bar R}_{{\rm{H}},v}+ {\bar R}_{{\rm{H}},W} $, respectively. The transmit power of the HD-relay and FD-relay are set to $p_d=(p-10)$ dBm. We can see that the network throughput gap between the RIS-aided NOMA network and the other pair of relay aided networks becomes smaller, when the number of RISs is increased. Observe that for the case of $N=18$, the proposed RIS-aided NOMA network is capable of outperforming two relay aided networks, which indicates that the RIS-aided NOMA network becomes more competitive, when the number of RISs is high enough.

\begin{figure}[t!]
\centering
\includegraphics[width =3 in]{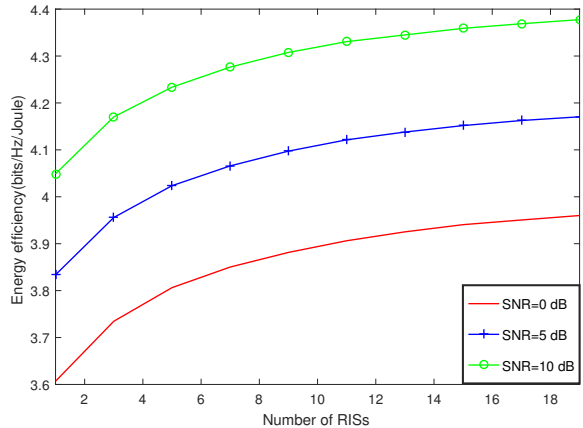}
\caption{EE of the proposed RIS-aided NOMA network versus the number of RISs at different SNR. The amplifier efficiency at the BS is set to ${\varepsilon _b}$ =1.2. The static power at the BS and users are set to $P_{\rm{B,s}}=9$ dBW as well as $P_{\rm{U}}=10$ dBm, respectively. The energy consumption of each RIS is set to $P_L=10$ dBm.}
\label{EE fig}
\vspace{-0.2in}
\end{figure}

\emph{8) Energy Efficiency:} Fig.~\ref{EE fig} evaluates the EE of the proposed RIS-aided NOMA network versus the number of RISs. On the one hand, it is can be observed that the EE improves as the number of RISs increases. However, observe that the slope of the EE curve is decreasing, which indicates that there exists an optimal value of the number of RISs that maximizes the EE.
Furthermore, in contrast to the conventional relay networks~\cite{renzo_RIS_relay}, it is worth noting that the EE can be increased upon increasing the transmit power at the BS.

\section{Conclusions}

In this article, we first reviewed the recent advances in RIS-aided networks. In order to illustrate the impact of RISs, we adopted a SISO network, where the passive beamforming weights of the RISs were designed. Both the best-case and worst-case of our new channel statistics, OPs, ergodic rates, EEs and SEs were derived in closed-form for characterizing the system performance.
In NG networks, an important future direction is to extend the proposed model to RIS-aided MIMO-NOMA network, where the active beamforming and passive beamforming, as well as detection vectors can be jointly designed for improving the network's throughput.

\numberwithin{equation}{section}
\section*{Appendix~A: Proof of Lemma~\ref{lemma1:new state of effective channel gain}} \label{Appendix:As}
\renewcommand{\theequation}{A.\arabic{equation}}
\setcounter{equation}{0}
In wireless communications, a pair of wave sources may be deemed perfectly coherent if they only have a constant phase difference but the same frequency, as well as the same waveform. In this context, a RIS array exhibit spatial coherence because its elements at the opposite ends of the array have a fixed phase relationship.
Therefore, there are two potential scenarios for modeling the channel gain of RIS-aided networks: 1) When the length of the BS-user link and BS-RIS-user link is nearly identical, the received signals can be coherent by considered, and thus the channel gain is given by
\begin{equation}\label{maximum achievable gain_appen}
{\left| {{{\tilde h}_W}} \right|^2} ={\left| {{{\bf{g}}_W}{\bf{\Phi h}}d_1^{ - \frac{{{\alpha _l}}}{2}}d_{2,W}^{ - \frac{{{\alpha _l}}}{2}} + {r_W}d_{3,W}^{ - \frac{{{\alpha _n}}}{2}}} \right|} ^2.
\end{equation}
2) On the other hand, since in practice, the length of the BS-user and BS-RIS-user links usually varies substantially, the received signals cannot be considered as coherent waves~\cite{Coherence_wave}.
Thus, when the reflected BS-RIS-user and the direct BS-user signals are co-phased, the BS-user and the BS-RIS-user signals can be boosted at the prioritized user $W$ by utilizing the classic maximum ratio combining (MRC) technique having the following channel gain:
\begin{equation}\label{effective channel gain after design appen A}
{\left| {{{\tilde h}_W}} \right|^2} = {\left| {{{\bf{g}}_W}{\bf{\Phi h}}} \right|^2}{\left( {{d_1}{d_{2,W}}} \right)^{ - {\alpha _l}}} + {\left| {{r_W}} \right|^2}d_{3,W}^{ - {\alpha _n}}.
\end{equation}
Note that for the first scenario, where the direct BS-user and the reflected BS-RIS-user links are coherent waves, the results of~\eqref{maximum achievable gain_appen} can be transformed into~\eqref{effective channel gain after design appen A}, provided that the number of RISs is high enough. Thus, in the rest of this article, we only analyze the network's performance for the second scenario.

Based on the passive beamforming design in~\eqref{define of the passive beamforming}, the effective channel gain of the prioritized user $W$ can be written as
\begin{equation}\label{Appendix A after RIS}
{\left| {{{\tilde h}_W}} \right|^2} = {\left( {\sum\limits_{n = 1}^N {\left| {{g_{W,n}}{h_n}} \right|} } \right)^2}{\left( {{d_1}{d_{2,W}}} \right)^{ - {\alpha _l}}} + {\left| {{r_W}} \right|^2}d_{3,W}^{ - {\alpha _n}}.
\end{equation}

By exploiting the fact that the elements of ${\left| {{{\rm \bf{g}}_{W}}} \right|}$ and ${\left| {{{\rm \bf{h}}}} \right|}$, as well as of ${\left| {{{{r}}_{W}}} \right|}$ are i.i.d., the worst-case of effective channel gain matrix can be transformed into

\begin{equation}\label{Apeendix effective channel gain}
\begin{aligned}
{\left| {{{{\bf{\tilde h}}}_W}} \right|^2} &= \sum\limits_{n = 1}^N {{{\left| {{g_{W,n}}{h_n}} \right|}^2}} {\left( {{d_1}{d_{2,W}}} \right)^{ - {\alpha _l}}} + {\left| {{r_W}} \right|^2}d_{3,W}^{ - {\alpha _n}} \\
&  = \sum\limits_{n = 1}^N {{{\left| {{g_{W,n}}} \right|}^2}{{\left| {{h_n}} \right|}^2}} {\left( {{d_1}{d_{2,W}}} \right)^{ - {\alpha _l}}} + {\left| {{r_W}} \right|^2}d_{3,W}^{ - {\alpha _n}}.
\end{aligned}
\end{equation}

Note that the elements of the channel matrix ${\left| {{{\rm \bf{g}}_{W}}} \right|}$ and ${\left| {{{\rm \bf{h}}}} \right|}$ obey the Nakagami distribution having the fading parameters $m_W$ and $m_1$, respectively. By exploiting the property of random variables, we obtain the mean and variance as follows
\begin{equation}\label{appendix first mean}
{E_1} = {\mathbb E}\left( { {{{{\left| {g_{W,n}} \right|}^2}} } } \right) {\mathbb E}\left( {{{\left| {{{{h}}_{n}}} \right|}^2}} \right) = 1,
\end{equation}
and
\begin{equation}\label{appendix first variance}
\begin{aligned}
& {V_1} = \left( { {\mathbb E} {{\left( { {{{\left| {g_{W,n}} \right|}^2}} } \right)}^2} + V\left( { {{{\left| {g_{W,n}} \right|}^2}} } \right)} \right)\left( { {\mathbb E}{{\left( {{{\left| {{{{h}}_{n}}} \right|}^2}} \right)}^2} + V\left( {{{\left| {{{{h}}_{n}}} \right|}^2}} \right)} \right) \\
& -  {\mathbb E}{\left( {{{\left| {g_{W,n}} \right|}^2}} \right)^2} { {\mathbb E} {{\left( { {{{\left| {{{h}}_{n}} \right|}^2}} } \right)}^2}}
 = \frac{{\left( {1 + {m_1} + {m_W}} \right)}}{{{m_1}{m_W}}}.
\end{aligned}
\end{equation}
Thus, the distribution can be written as
\begin{equation}\label{appendix channel multiple distribution}
{{{\left| {{g_{W,n}}{h_{n}}} \right|}^2}} \sim \mathcal{CN}\left( {1,{m_l}} \right),
\end{equation}
where $m_l=\frac{1+m_1+m_W}{m_1 {m_W}} $, and thereby the mean and variance of the effective channel gain is given by
\begin{equation}\label{appendix effective channel distribution}
\sum\limits_{n = 1}^N {{{\left| {{g_{W,n}}{h_n}} \right|}^2}} {\left( {{d_1}{d_{2,W}}} \right)^{ - {\alpha _l}}}\sim {\mathcal{CN}} \left( {N{{\left( {{d_1}{d_{2,W}}} \right)}^{ - {\alpha _l}}},N{m_l}{{\left( {{d_1}{d_{2,W}}} \right)}^{ - 2{\alpha _l}}}} \right).
\end{equation}

Due to the fact that user $W$ also detects the signal transmitted from the direct BS-user link, the worst-case of the effective channel gain of user $W$ can be rewritten as
\begin{equation}\label{appendix effective channel distribution}
{\left| {{{{\rm{\tilde h}}}_{W,l}}} \right|^2}\sim {\mathcal{CN}}\left( {N{{\left( {{d_1}{d_{2,W}}} \right)}^{ - {\alpha _l}}} + d_{3,W}^{ - {\alpha _n}},N{m_l}{{\left( {{d_1}{d_{2,W}}} \right)}^{ - 2{\alpha _l}}} + d_{3,W}^{ - 2{\alpha _n}}} \right).
\end{equation}
After some algebraic manipulations, we obtain the effective channel gain in a more elegant form in~\eqref{New Gamma distribution in Lemma_w-th user}.

We then turn our attention to the best-case of the effective channel gain of user $W$. 
By exploiting Cauchy-Schwarz inequality~\cite{ZHAO_Cauchy_schwarz}, the effective channel gain can be written as
\begin{equation}\label{Appendix upper bound approximation}
\left( \sum\limits_{n = 1}^N {{{\left| {{g_{W,n}}{h_n}} \right|}}} \right)^2
 \le  \left( \sum \limits_{n = 1}^N {{{ { \left| {g_{W,n}} \right| ^2} }}} \right) \left( \sum \limits_{n = 1}^N {{{ {\left|  { {h_{n}}}\right|^2 } }}} \right).
\end{equation}
Similar to the procedures in~\eqref{appendix first mean} to~\eqref{appendix effective channel distribution}, the best-case of the distribution can be obtained as follows:
\begin{equation}\label{appendix effective channel distribution upper bound}
{\left| {{{{\rm{\tilde h}}}_{W,l}}} \right|^2}\sim {\mathcal{CN}}\left( {{N^2}{{\left( {{d_1}{d_{2,W}}} \right)}^{ - {\alpha _l}}} + d_{3,W}^{ - {\alpha _n}},{N^2}{m_u}{{\left( {{d_1}{d_{2,W}}} \right)}^{ - 2{\alpha _l}}} + d_{3,W}^{ - 2{\alpha _n}}} \right).
\end{equation}
Hence, the best-case distribution can be derived in terms of~\eqref{New Gamma distribution in Lemma_w-th user upper bound}, and the proof is complete.

\numberwithin{equation}{section}
\section*{Appendix~B: Proof of Theorem~\ref{Theorem1:Outage W user closed form by incomlete gamma}} \label{Appendix:Bs}
\renewcommand{\theequation}{B.\arabic{equation}}
\setcounter{equation}{0}

Let us first consider the worst-case of the prioritized user $W$. Based on the OP defined in~\eqref{Outage Defination}, the worst-case of the OP can be rewritten as
\begin{equation}\label{appendix B outage defination}
{P_W} = \mathbb{ P} \left( {{{\left| {{{\tilde h}_{W,l}}} \right|}^2} < {I_v}} \right) + \mathbb{ P} \left( {{I_v} < {{\left| {{{\tilde h}_{W,l}}} \right|}^2} < {I_W}} \right).
\end{equation}

Since the users are ordered based on their effective channel gain, the marginal PDF of user $W$ is given by~\cite{Order_Statistics}
\begin{equation}\label{appendix B marginal PDF}
{f_W}(x) = {W}{{\tilde f}_W}(x){\left( {{{\tilde F}_W}(x)} \right)^{W - 1}},
\end{equation}
where ${{\tilde f}_W}(x)$ and ${{\tilde F}_W}(x)$ represent the PDF and cumulative density function (CDF) of the unordered effective channel gain associated with ${{\tilde f}_W}(x) = \frac{{{x^{{k_1} - 1}}}}{{\Gamma ({k_1})\lambda _1^{{k_1}}}}\exp ( - \frac{x}{\lambda_1 })$ and ${{\tilde F}_W}(x) = 1 - \frac{{\gamma \left( {{k_1},\frac{x}{\lambda_1 }} \right)}}{{\Gamma ({k_1})}}$.

Based on the marginal PDF given in~\eqref{appendix B marginal PDF}, the OP is given by
\begin{equation}\label{appendix B outage expression}
{P_W} = {W}\int\limits_0^{{I_{W*}}} {{\tilde f}_W}(x){{\left( {{{\tilde F}_W}(x)} \right)}^{W - 1}}dx.
\end{equation}
After some algebraic manipulations, the OP of the prioritized user $W$ in~\eqref{outage analytical results W in theorem1} can be obtained. The proof is complete.

\numberwithin{equation}{section}
\section*{Appendix~C: Proof of Corollary~\ref{corollary1:Outage v user asymptotic}} \label{Appendix:Cs}
\renewcommand{\theequation}{C.\arabic{equation}}
\setcounter{equation}{0}

In order to glean further engineering insights, we first expand the lower incomplete Gamma function as follows~\cite{Table_of_integrals}:
\begin{equation}\label{Appendix C Lower incomplete gamma expansion}
\begin{aligned}
\gamma \left( {{k_1},\frac{{{I_{W*}}}}{{{\lambda _1}}}} \right) = & \sum\limits_{s = 0}^\infty  {} \frac{{\Gamma \left( {{k_1}} \right)}}{{\Gamma \left( {{k_1} + s + 1} \right)}} {\left( {\frac{{{I_{W*}}}}{{{\lambda _1}}}} \right)^{{k_1} + s}}\exp \left( -{\frac{{{I_{W*}}}}{{{\lambda _1}}}} \right).
\end{aligned}
\end{equation}

In the high-SNR regime, recall that $\mathop {\lim }\limits_{x \to 0 } \left( {1 - {e^{ - x}}} \right) \approx x$. Hence the OP of user $W$ can be approximated in the high-SNR regime as follows:
\begin{equation}\label{Apeendix C first approx}
{{\bar P}_W} ={\left( {\sum\limits_{s = 0}^\infty  {{\frac{1}{{\Gamma \left( {{k_1} + s + 1} \right)}}}} {{\left( {\frac{{{I_{W*}}}}{{{\lambda _1}}}} \right)}^{{k_1} + s}}\left( {1 - \frac{{{I_{W*}}}}{{{\lambda _1}}}} \right)} \right)^{W}}.
\end{equation}
Upon involving the binomial expansion and after some algebraic manipulations, the approximate result can be further transformed into
\begin{equation}\label{Appendix C second transform}
\begin{aligned}
 {{\bar P}_W} &= {\left( {\sum\limits_{s = 0}^\infty  {\frac{1}{{\Gamma \left( {{k_1} + s + 1} \right)}}} {{\left( {\frac{{{I_{W*}}}}{{{\lambda _1}}}} \right)}^s}} \right)^W} \sum\limits_{j = 0}^W {\left( {\begin{array}{*{20}{c}}
W\\
j
\end{array}} \right){{( - 1)}^j}} {\left( {\frac{{{I_{W*}}}}{{{\lambda _1}}}} \right)^{{k_1}W + j}}.
\end{aligned}
\end{equation}
Thus, after some algebraic manipulations, the results in~\eqref{asymptotic result v user in corollary1} can be obtained, and the proof is complete.


\numberwithin{equation}{section}
\section*{Appendix~D: Proof of Theorem~\ref{theorem3:ergodic rate W-th user}} \label{Appendix:Ds}
\renewcommand{\theequation}{D.\arabic{equation}}
\setcounter{equation}{0}

Let us commence by expressing the worst-case on the ergodic rate of the prioritized user $W$ as follows:
\begin{equation}\label{Appendix D first define}
\begin{aligned}
{R_{W,l}} & = \mathbb{E} \left\{ {{{\log }_2}\left[ {1 + SIN{R_W}\left( x \right)} \right]} \right\} =  - \int\limits_0^\infty  {{{\log }_2}(1 + x)} d\left[ {1 - {{F}}\left( x \right)} \right] \\
& = \frac{1}{{\ln \left( 2 \right)}}\int\limits_0^\infty  {\frac{{1 - F\left( x \right)}}{{1 + x}}} dx.
\end{aligned}
\end{equation}

The CDF of user $W$ can be calculated as
\begin{equation}\label{Appendix D CDF define}
F\left( x \right) = {\left( {\frac{{\gamma \left( {{k_1}, Cx } \right)}}{{\Gamma ({k_1})}}} \right)^{W}}.
\end{equation}

In order to derive the closed-form expression of the worst-case, we first round the shape parameter to the closest integer, i.e., $\bar k = \left[ {{k_1}} \right]$. Hence, the lower incomplete Gamma function can be further expanded to
\begin{equation}\label{Appendix D lower incomplete gamma function expansion}
\frac{{\gamma \left( {\bar k,Cx} \right)}}{{\Gamma (\bar k)}}{\rm{ = }}\left( {1{\rm{ - }}\sum\limits_{i = 0}^{\bar k - 1} {\frac{{{{\left( {Cx} \right)}^i}}}{{i!}}{e^{ - Cx}}} } \right).
\end{equation}

By utilizing the binomial expansion, the result in~\eqref{Appendix D lower incomplete gamma function expansion} can be further transformed into
\begin{equation}\label{Apeendix D binomial expansion}
\begin{aligned}
& {\left( {1{\rm{ - }}\sum\limits_{i = 0}^{\bar k - 1} {\frac{{{{\left( {Cx} \right)}^i}}}{{i!}}{e^{ - Cx}}} } \right)^W} = \sum\limits_{s = 0}^W { \begin{pmatrix}
W\\
s
\end{pmatrix} } {( - 1)^s}{\left( {\sum\limits_{i = 0}^{\bar k - 1} {\frac{{{{\left( {Cx} \right)}^i}}}{{i!}}{e^{ - Cx}}} } \right)^s}.
\end{aligned}
\end{equation}
Thus, the ergodic rate can be written as
\begin{equation}\label{Appendic D ergodic rate before integral}
\begin{aligned}
& {R_{W,l}} =  \frac{1}{{\ln \left( 2 \right)}}\sum\limits_{s = 1}^W { \begin{pmatrix}
W\\
s
\end{pmatrix} } {( - 1)^s}\int\limits_0^\infty  {\frac{{{e^{ - Csx}}{{\left( {\sum\limits_{i = 0}^{\bar k - 1} {\frac{{{{\left( {Cx} \right)}^i}}}{{i!}}} } \right)}^s}}}{{1 + x}}} dx.
\end{aligned}
\end{equation}

We then expand~\eqref{Appendic D ergodic rate before integral} by using the multi-nomial theorem as follows~\cite{Mathematics}:
\begin{equation}\label{Appendix D multinomial expansion}
\begin{aligned}
& {\left( {\sum\limits_{i = 0}^{\bar k - 1} {\frac{{{{\left( {Cx} \right)}^i}}}{{i!}}{e^{ - Cx}}} } \right)^s} = \sum\limits_{{a_1} +  \ldots  + {a_{\bar k}} = s} {}  \begin{pmatrix}
s\\
{a_1}, \ldots ,{a_{\bar k}}
\end{pmatrix} {\prod\limits_{t = 1}^{\bar k} {\left( {\frac{{{{\left( {Cx} \right)}^{t - 1}}}}{{(t - 1)!}}} \right)} ^{{a_t}}},
\end{aligned}
\end{equation}
where $ \begin{pmatrix}
s\\
{a_1}, \ldots ,{a_{\bar k}}
\end{pmatrix}  = \frac{{s!}}{{{a_1}! \cdot  \ldots  \cdot {a_{\bar k}}!}}$. Thus, the result can be rewritten as
\begin{equation}\label{Appendix D before final}
\begin{aligned}
{R_{W,l}} &= \sum\limits_{s = 0}^W { \begin{pmatrix}
W\\
s
\end{pmatrix} } {( - 1)^s}\frac{1}{{\bar k}} \sum\limits_{{a_1} +  \ldots  + {a_{\bar k}} = s} {} \begin{pmatrix}
s\\
{a_1}, \ldots ,{a_{\bar k}}
\end{pmatrix} {\prod\limits_{t = 1}^{\bar k} {\left( {\frac{{{{\left( C \right)}^{t - 1}}}}{{(t - 1)!}}} \right)} ^{{a_t}}}  \\
&\times \int\limits_0^\infty  {\frac{{{x^{\left( {t - 1} \right){a_t}}}\exp \left( { - Csx} \right)}}{{1 + x}}} dx.
\end{aligned}
\end{equation}

Hence, the tractable approximate results can be derived as
\begin{equation}\label{appendix D final result}
\begin{aligned}
& {R_{W,l}}  = \sum\limits_{s = 0}^W { \begin{pmatrix}
W\\
s
\end{pmatrix} } {( - 1)^s}\frac{1}{{\bar k}} \sum\limits_{{a_1} +  \ldots  + {a_{\bar k}} = s} {} \begin{pmatrix}
s\\
{a_1}, \ldots ,{a_{\bar k}}
\end{pmatrix} {\prod\limits_{t = 1}^{\bar k} {\left( {\frac{{{{\left( C \right)}^{t - 1}}}}{{(t - 1)!}}} \right)} ^{{a_t}}}  \\
& \times \left( {\exp (Cs)Ei( - Cs) + \sum\limits_{i = 1}^{\left( {t - 1} \right){a_t}} {{{( - 1)}^{i - 1}}(i - 1)!{{(Cs)}^i}} } \right).
\end{aligned}
\end{equation}
Thus, the worst-case on the ergodic rate of user $W$ is obtained in~\eqref{asympto W-th erogodic rate in corollary4}, and the proof is complete.

\begin{spacing}{1.25}
\bibliographystyle{IEEEtran}
\bibliography{IEEEabrv,NOMA_RIS}

\begin{thebibliography}{10}
\providecommand{\url}[1]{#1}
\csname url@samestyle\endcsname
\providecommand{\newblock}{\relax}
\providecommand{\bibinfo}[2]{#2}
\providecommand{\BIBentrySTDinterwordspacing}{\spaceskip=0pt\relax}
\providecommand{\BIBentryALTinterwordstretchfactor}{4}
\providecommand{\BIBentryALTinterwordspacing}{\spaceskip=\fontdimen2\font plus
\BIBentryALTinterwordstretchfactor\fontdimen3\font minus
  \fontdimen4\font\relax}
\providecommand{\BIBforeignlanguage}[2]{{%
\expandafter\ifx\csname l@#1\endcsname\relax
\typeout{** WARNING: IEEEtran.bst: No hyphenation pattern has been}%
\typeout{** loaded for the language `#1'. Using the pattern for}%
\typeout{** the default language instead.}%
\else
\language=\csname l@#1\endcsname
\fi
#2}}
\providecommand{\BIBdecl}{\relax}
\BIBdecl

\bibitem{5G_NR}
S.~{Lien}, S.~{Shieh}, Y.~{Huang}, B.~{Su}, Y.~{Hsu}, and H.~{Wei}, ``5{G} new
  radio: Waveform, frame structure, multiple access, and initial access,''
  \emph{IEEE Commun. Mag.}, vol.~55, no.~6, pp. 64--71, Jun. 2017.

\bibitem{5G_NR_2}
S.~{Parkvall}, E.~{Dahlman}, A.~{Furuskar}, and M.~{Frenne}, ``{NR}: The new
  5{G} radio access technology,'' \emph{IEEE Commun. Standards Mag.}, vol.~1,
  no.~4, pp. 24--30, Dec. 2017.

\bibitem{LIS_zhangjiayi_mag}
J.~Zhang, E.~Björnson, M.~Matthaiou, D.~W.~K. Ng, H.~Yang, and D.~J. Love,
  ``Multiple antenna technologies for beyond 5{G},'' \emph{Arxiv}, vol.
  1910.00092, Sep. 2019.

\bibitem{LIS_smart}
Y.~Liang, R.~Long, Q.~Zhang, J.~Chen, H.~V. Cheng, and H.~Guo, ``Large
  intelligent surface/antennas ({LISA}): Making reflective radios smart,''
  \emph{J. Commun. Inf. Networks}, vol.~4, no.~2, pp. 40--50, Jun. 2019.

\bibitem{RIS_mag_basar}
E.~Basar, ``Transmission through large intelligent surfaces: A new frontier in
  wireless communications,'' \emph{Arxiv}, vol. 1902.08463v2, Apr. 2019.

\bibitem{NOMA_mag_Ding}
Z.~Ding, Y.~Liu, J.~Choi, Q.~Sun, M.~Elkashlan, C.~I, and H.~V. Poor,
  ``Application of non-orthogonal multiple access in {LTE} and 5{G} networks,''
  \emph{IEEE Commun. Mag.}, vol.~55, no.~2, pp. 185--191, Feb. 2017.

\bibitem{PairingDING2016}
Z.~Ding, P.~Fan, and H.~V. Poor, ``Impact of user pairing on 5{G} nonorthogonal
  multiple-access downlink transmissions,'' \emph{IEEE Trans. Veh. Technol.},
  vol.~65, no.~8, pp. 6010--6023, Aug. 2016.

\bibitem{Massive_NOMA_Cellular_IoT}
M.~{Shirvanimoghaddam}, M.~{Dohler}, and S.~J. {Johnson}, ``Massive
  non-orthogonal multiple access for cellular {IoT}: Potentials and
  limitations,'' \emph{IEEE Commun. Mag.}, vol.~55, no.~9, pp. 55--61, Sep.
  2017.

\bibitem{NOMA_5G_beyond_Liu}
Y.~Liu, Z.~Qin, M.~Elkashlan, Z.~Ding, A.~Nallanathan, and L.~Hanzo,
  ``Nonorthogonal multiple access for 5{G} and beyond,'' \emph{Proc. of the
  IEEE}, vol. 105, no.~12, pp. 2347--2381, Dec. 2017.

\bibitem{Islam_NOMA_survey}
S.~M.~R. Islam, N.~Avazov, O.~A. Dobre, and K.~Kwak, ``Power-domain
  non-orthogonal multiple access ({NOMA}) in 5{G} systems: Potentials and
  challenges,'' \emph{IEEE Commun. Surveys Tuts.}, vol.~19, no.~2, pp.
  721--742, Secondquarter 2017.

\bibitem{NOMA_large_heter}
Y.~{Liu}, Z.~{Qin}, M.~{Elkashlan}, A.~{Nallanathan}, and J.~A. {McCann},
  ``Non-orthogonal multiple access in large-scale heterogeneous networks,''
  \emph{IEEE J. Sel. Areas Commun.}, vol.~35, no.~12, pp. 2667--2680, Dec.
  2017.

\bibitem{LIS_magazine_multi_scenarios}
Q.~Wu and R.~Zhang, ``Towards smart and reconfigurable environment: Intelligent
  reflecting surface aided wireless network,'' \emph{Arxiv}, vol. 1905.00152v3,
  Jun. 2019.

\bibitem{reconfig_meta_surf_1}
M.~D. Renzo \emph{et~al.}, ``Smart radio environments empowered by {AI}
  reconfigurable meta-surfaces: An idea whose time has come,'' \emph{Arxiv},
  vol. 1903.08925, Mar. 2019.

\bibitem{reconfig_meta_surf_2}
M.~D. Renzo and J.~Song, ``Reflection probability in wireless networks with
  metasurface-coated environmental objects: An approach based on random spatial
  processes,'' \emph{Arxiv}, vol. 1901.01046v1, Jan. 2019.

\bibitem{LIS_compare_relay}
E.~Bjornson, O.~\"{O}zdogan, and E.~G. Larsson, ``Intelligent reflecting
  surface vs. decode-and-forward: How large surfaces are needed to beat
  relaying?'' \emph{Arxiv}, vol. 1906.03949, Jun. 2019.

\bibitem{glob_energy_model}
C.~{Huang}, G.~C. {Alexandropoulos}, A.~{Zappone}, M.~{Debbah}, and C.~{Yuen},
  ``Energy efficient multi-user {MISO} communication using low resolution large
  intelligent surfaces,'' in \emph{2018 {IEEE GLOBECOM} Workshops (GC Wkshps)},
  Dec. 2018, pp. 1--6.

\bibitem{energy_model_LIS}
C.~Huang, A.~Zappone, G.~Alexandropoulos, M.~Debbah, and C.~Yuen,
  ``Reconfigurable intelligent surfaces for energy efficiency in wireless
  communication,'' \emph{Arxiv}, vol. 1810.06934v5, Jun. 2019.

\bibitem{Renzo_PHY_security_confe}
H.~Han, J.~Zhao, D.~Niyato, M.~D. Renzo, and Q.~Pham, ``Intelligent reflecting
  surface aided network: Power control for physical-layer broadcasting,''
  \emph{arXiv}, vol. 1910.14383v1, pp. 1--1, Oct. 2019.

\bibitem{PLS_LIS_ZhangRui}
X.~Guan, Q.~Wu, and R.~Zhang, ``Intelligent reflecting surface assisted secrecy
  communication via joint beamforming and jamming,'' \emph{arXiv}, vol.
  1907.12839v3, pp. 1--1, Jul. 2019.

\bibitem{Swipt_LIS_ZhangRui}
Q.~Wu and R.~Zhang, ``Weighted sum power maximization for intelligent
  reflecting surface aided {SWIPT},'' \emph{arXiv}, vol. 1907.05558v2, pp.
  1--1, Jul. 2019.

\bibitem{Renzo_mmwave_signal_enhancement}
N.~S. Perović, M.~D. Renzo, and M.~F. Flanagan, ``Channel capacity
  optimization using reconfigurable intelligent surfaces in indoor mm{W}ave
  environments,'' \emph{arXiv}, vol. 1910.14310v1, pp. 1--1, Oct. 2019.

\bibitem{Lv_coverage_enhancement}
Y.~Cao and T.~Lv, ``Intelligent reflecting surface aided multi-user
  millimeter-wave communications for coverage enhancement,'' \emph{arXiv}, vol.
  1910.02398v1, pp. 1--1, Oct. 2019.

\bibitem{Zhou_MISO_multi_cluster}
G.~Zhou, C.~Pan, H.~Ren, K.~Wang, and A.~Nallanathan, ``Intelligent reflecting
  surface aided multigroup multicast {MISO} communication systems,''
  \emph{arXiv}, vol. 1909.04606v2, pp. 1--1, Sep. 2019.

\bibitem{ZhangRui_MISO_beams_discrete_2}
Q.~Wu and R.~Zhang, ``Beamforming optimization for wireless network aided by
  intelligent reflecting surface with discrete phase shifts,'' \emph{Arxiv},
  vol. 1906.03165v2, Jun. 2019.

\bibitem{shuowen_RIS}
S.~Zhang and R.~Zhang, ``Capacity characterization for intelligent reflecting
  surface aided {MIMO} communication,'' \emph{Arxiv}, vol. 1910.13636v1, Oct.
  2019.

\bibitem{Hou_RIS_MIMO_global_algrithm}
T.~Hou, Y.~Liu, Z.~Song, X.~Sun, Y.~Chen, and L.~Hanzo, ``{MIMO} assisted
  networks relying on large intelligent surfaces: A stochastic geometry
  model,'' \emph{arXiv}, vol. 1910.00959v1, pp. 1--1, Oct. 2019.

\bibitem{DING_RIS_NOMA_letter}
Z.~Ding and H.~V. Poor, ``A simple design of {IRS-NOMA} transmission,''
  \emph{arXiv}, vol. 1907.09918v2, pp. 1--1, Jul. 2019.

\bibitem{yuanwei_NOMA_RIS}
X.~Mu, Y.~Liu, L.~Guo, J.~Lin, and N.~Al-Dhahir, ``Exploiting intelligent
  reflecting surfaces in multi-antenna aided {NOMA} systems,'' \emph{Arxiv},
  vol. 1910.13636v1, Oct. 2019.

\bibitem{MISO_with_directlink}
X.~Yu, D.~Xu, and R.~Schober, ``{MISO} wireless communication systems via
  intelligent reflecting surfaces,'' \emph{Arxiv}, vol. 1904.12199v1, Apr.
  2019.

\bibitem{NOMA_RIS_Fu}
M.~Fu, Y.~Zhou, and Y.~Shi, ``Intelligent reflecting surface for downlink
  non-orthogonal multiple access networks,'' \emph{arXiv}, vol. 1906.09434v3,
  pp. 1--1, Jun. 2019.

\bibitem{LIS_perform_Anal}
E.~Basar, ``Large intelligent surface-based index modulation: A new beyond
  {MIMO} paradigm for 6{G},'' \emph{Arxiv}, vol. 1904.06704v1, Apr. 2019.

\bibitem{RIS_NOMA_Rice}
G.~Yang, X.~Xu, and Y.~Liang, ``Intelligent reflecting surface assisted
  non-orthogonal multiple access,'' \emph{Arxiv}, vol. 1907.03133v1, Jul. 2019.

\bibitem{wireless_communication_goldsmith}
A.~Goldsmith, \emph{Wireless Communication}.\hskip 1em plus 0.5em minus
  0.4em\relax Cambridge University Press, 2nd ed, 2010.

\bibitem{Yi_anlog_beam}
W.~{Yi}, Y.~{Liu}, A.~{Nallanathan}, and M.~{Elkashlan}, ``Clustered
  millimeter-wave networks with non-orthogonal multiple access,'' \emph{IEEE
  Trans. Commun.}, vol.~67, no.~6, pp. 4350--4364, Jun. 2019.

\bibitem{Hou_naka_order}
T.~{Hou}, X.~{Sun}, and Z.~{Song}, ``Outage performance for non-orthogonal
  multiple access with fixed power allocation over {N}akagami-${m}$ fading
  channels,'' \emph{IEEE Commun. Lett.}, vol.~22, no.~4, pp. 744--747, Apr.
  2018.

\bibitem{Hou_Single_UAV}
T.~{Hou}, Y.~{Liu}, Z.~{Song}, X.~{Sun}, and Y.~{Chen}, ``Multiple antenna
  aided {NOMA} in {UAV} networks: A stochastic geometry approach,'' \emph{IEEE
  Trans. Commun.}, vol.~67, no.~2, pp. 1031--1044, Feb. 2019.

\bibitem{EE_model_massive_MIMO}
L.~N. {Ribeiro}, S.~{Schwarz}, M.~{Rupp}, and A.~L.~F. {de Almeida}, ``Energy
  efficiency of mm{W}ave massive {MIMO} precoding with low-resolution {DAC}s,''
  \emph{IEEE J. Sel. Topics Signal Process}, vol.~12, no.~2, pp. 298--312, May
  2018.

\bibitem{yuanwei_cooperative}
Y.~{Liu}, Z.~{Ding}, M.~{Elkashlan}, and H.~V. {Poor}, ``Cooperative
  non-orthogonal multiple access with simultaneous wireless information and
  power transfer,'' \emph{IEEE J. Sel. Areas Commun.}, vol.~34, no.~4, pp.
  938--953, Apr. 2016.

\bibitem{DF_relaying_outage}
J.~N. {Laneman}, D.~N.~C. {Tse}, and G.~W. {Wornell}, ``Cooperative diversity
  in wireless networks: Efficient protocols and outage behavior,'' \emph{IEEE
  Trans. Information Theory}, vol.~50, no.~12, pp. 3062--3080, Dec. 2004.

\bibitem{cooperative_Yue}
X.~{Yue}, Y.~{Liu}, S.~{Kang}, A.~{Nallanathan}, and Z.~{Ding}, ``Spatially
  random relay selection for full/half-duplex cooperative {NOMA} networks,''
  \emph{IEEE Trans. Commun.}, vol.~66, no.~8, pp. 3294--3308, Aug. 2018.

\bibitem{renzo_RIS_relay}
K.~Ntontin, J.~Song, and M.~D. Renzo, ``Multi-antenna relaying and
  reconfigurable intelligent surfaces: End-to-end {SNR} and achievable rate,''
  \emph{Arxiv}, vol. 1908.07967v2, pp. 1--1, Aug. 2019.

\bibitem{Coherence_wave}
R.~G. Winter and A.~M. Steinberg, \emph{Coherence}.\hskip 1em plus 0.5em minus
  0.4em\relax Access Science. McGraw-Hill., 2008.

\bibitem{ZHAO_Cauchy_schwarz}
Z.~Zhao, Z.~Ding, M.~Peng, and Y.~Li, ``A full-cooperative diversity
  beamformingscheme in two-way amplify-and-forward relay systems,''
  \emph{Digital Communications and Networks}, vol.~1, no.~1, pp. 57--67, Mar.
  2015.

\bibitem{Order_Statistics}
H.~N. Nagaraja and H.~A. David, \emph{Order Statistics}.\hskip 1em plus 0.5em
  minus 0.4em\relax John Wiley, New York, 3rd ed, 2003.

\bibitem{Table_of_integrals}
I.~S. Gradshteyn and I.~M. Ryzhik, \emph{Table of Integrals, Series and
  Products}.\hskip 1em plus 0.5em minus 0.4em\relax New York: Academic Press,
  6th ed, 2000.

\bibitem{Mathematics}
N.~Bourbaki, \emph{Elements of the History of Mathematics Paperback}.\hskip 1em
  plus 0.5em minus 0.4em\relax Springer Berlin Heidelberg, 2nd ed, 2008.

\end{thebibliography}
\end{spacing}
\end{document}